\documentstyle[11pt]{article}
\def\Tr{\,{\rm Tr}\,}
\begin{document}
\def\beq{\begin{equation}}
\def\eeq{\end{equation}}
\def\beqa{\begin{eqnarray}}
\def\eeqa{\end{eqnarray}}
\begin{titlepage}
\vspace*{-1cm}
\noindent
\phantom{bla}
\hfill{UMHEP-414}
\\
\vskip 2.5cm
\begin{center}
{\Large\bf Two-loop Analysis of Vector Current Propagators
in Chiral Perturbation Theory}
\end{center}
\vskip 1.5cm
\begin{center}
{\large Eugene Golowich and Joachim Kambor
\footnote{address after September 1, 1994: Division de Physique Th\'eorique,
Institut de Physique Nucl\'eaire, F-91406 Orsay Cedex, France} }
\\
\vskip .3cm
Department of Physics and Astronomy \\
University of Massachusetts \\
Amherst MA 01003 USA\\
\vskip .3cm
\end{center}
\vskip 2cm
\begin{abstract}
\noindent
We perform a calculation of the isospin and hypercharge
vector current propagators ($\Delta_{{\rm V}33}^{\mu\nu}(q^2)$
and $\Delta_{{\rm V}88}^{\mu\nu}(q^2)$) to two loops in chiral
perturbation theory.  The analysis is carried out with
straightforward Feynman diagram methods by making appropriate use
of external vector sources.  Counterterms from the ${\cal O}(q^6)$
chiral lagrangian, required to absorb divergences and scale
dependence encountered at the two-loop level, are constructed.
Our final results are finite, covariant, and scale-independent.
Several applications are described, including a comparison of
the two-loop isospin vector spectral function with data and the
construction of new chiral sum rules.
\end{abstract}
\vfill
\end{titlepage}
\vskip2truecm
\section{\bf Introduction}
A decade ago, Gasser and Leutwyler performed a general one-loop analysis
of chiral perturbation theory.$^{\cite{gl1},\cite{gl2}}$   Among their
results were representations of the isospin vector current and axialvector
current propagators, valid up to ${\cal O}(q^4)$ corrections.
In this paper, we shall describe a calculation to two-loop order in
chiral perturbation theory\footnote{We stress for clarity's sake that our
analysis throughout will be in the standard ChPT and not the so-called
generalized ChPT of Ref.~\cite{nogen}.} (hereafter ChPT) of both the
isospin and hypercharge vector current propagators, yielding
representations valid up to ${\cal O}(q^6)$ corrections.

In the years following the publication of the Gasser-Leutwyler papers
cited above, there appeared a dramatic increase in the number of studies
utilizing ChPT to analyze hadrons at low energies.
However, most of these involved either tree-level or one-loop
calculations.  In fact, there currently exists a need for higher-order
calculations which meaningfully probe the convergence properties
of ChPT.  This forms an important impetus for performing
our calculation.

There is a more phenomenologically-oriented way of stating our
motivation which actually bears on future applications of our work --
the subject of chiral sum rules.  It has long been known that among the
predictions of chiral symmetry are the following set of sum rules,
\vspace{0.15cm}
\begin{eqnarray}
& & \int_{0}^\infty ds ~{\rho_{\rm V}(s) - \rho_{\rm A}(s)\over s}
= -4 L_{10}^r (M_K^2) + {1\over 48\pi^2} \ln{M_K^2 \over M_\pi^2}
- {1\over 32\pi^2} \ , \label{I0} \\
& & \int_{0}^\infty ds ~(\rho_{\rm V}(s) - \rho_{\rm A}(s))
= F_\pi^2 \ ,\label{I1} \\
& & \int_{0}^\infty ds \ s~(\rho_{\rm V}(s)
- \rho_{\rm A}(s)) = 0 \ \ , \label{I2} \\
& & \int_{0}^\infty ds ~s\ln\left({s\over\Lambda^2}\right)~
(\rho_{\rm V}(s) - \rho_{\rm A}(s)) =
-{16\pi^2 F_\pi^2\over 3e^2}  (m^2_{\pi^\pm} - m^2_{\pi^0} ) \ .
\label{I3}
\end{eqnarray}
\vspace{0.15cm}
In the first of these$^{\cite{kfsr}}$, the quantity $L_{10}^r (M_K^2)$
is the renormalized coefficient of an ${\cal O}(E^4)$ counterterm
operator appearing in the $SU(3)$ chiral lagrangian of $QCD$.$^{\cite{gl2}}$
We do not comment further on this quantity here insofar as much of
Sect.~4 and Sect.~5 is devoted to a study of counterterms and their
renormalization.  The next two relations are respectively
the first and second Weinberg sum rules.$^{\cite{w}}$  The final
sum rule is a formula for the $\pi^\pm$-$\pi^0$ mass splitting
in the chiral limit.$^{\cite{dgmly}}$  Although apparently containing
an arbitrary energy scale $\Lambda$, this sum rule is actually
independent of $\Lambda$ by virtue of the preceding relation.

Although the above sum rules were derived almost $30$ years ago,
a full phenomenological test of their content was not carried out
until recently$^{\cite{dg1}}$.  In principle, the underlying strategy
employed in Ref.~{\cite{dg1}} was to divide the energy region into
three intervals as follows.  In the large energy region,
from $s \to \infty$ down to energies where data first become available,
operator product expansion methods were used.  At intermediate
energies, the vector and axialvector spectral functions $\rho_{\rm V}(s)$
and $\rho_{\rm A}(s)$ were extracted
from tau decay distributions and $e^+ e^- \to 2\pi,4\pi$ cross sections.
Finally, the threshold behavior ($s \to 4m_\pi^2$) of the spectral
functions was constrained by means of chiral symmetry.
In practice, however, this final step was not found to be crucial
to the calculation and as a result was not implemented.
The analysis of Ref.~\cite{dg1} ultimately led to a successful fit to
all four chiral sum rules, which is highly nontrivial in view of
the strong distortions experienced by the spectral functions
from the various moments.

Let us study the threshold prediction of ChPT for the
isospin spectral function.  The dashed curve in Fig.~1 is the one-loop
prediction of Gasser and Leutwyler and is to be compared with
data taken from $e^+ e^- \to \pi^+ \pi^-$ scattering.  One readily sees
that it does not take long for the one-loop
prediction to disagree with experiment as the energy is increased
above the two-pion threshold.  This is one of several reasons
for computing to the next order of ChPT.$^{\cite{ff}}$

We conclude this section with an outline of the rest of the paper.
Section 2 contains a discussion of our overall strategy for
carrying out the calculation.  The lowest-order lagrangian
${\cal L}^{(2)}$ is used to illustrate certain aspects of the discussion.
Section 3 summarizes both the one-loop and two-loop chiral
analysis and introduces ${\cal L}^{(4)}$, the ${\cal O}(q^4)$
chiral lagrangian, to the calculation.  Section 4 covers the counterterm
analysis, reviewing the presence of the ${\cal O}(q^4)$ counterterms and
especially generating a complete list of ${\cal O}(q^6)$ counterterms
relevant for our purposes.
Section 5 contains a statement of our renormalization procedure.  The use
of the renormalization group (RG hereafter) to determine the scale
dependence of various counterterms is discussed and the RG equations
are solved to the needed order.  Section 6 contains the explicit
representations for the isospin and hypercharge vector current
propagators which result from our calculation. Section 7 contains two
applications of our results. The improved isospin vector spectral
function, now containing next-to-leading order contributions, is again
compared with data. In addition, the asymptotic behaviour of the
vector-propagators is employed to derive a new set of chiral sum rules.
We conclude the main body of the paper in Section 8 by summarizing our
findings and proposing further research on the subject.  There are also
two Appendices, the first devoted to defining the $d$-dimensional integrals
which occur in our calculation and the second to giving an independent
determination of the isospin spectral function $\rho_{{\rm V}33} (s)$
at next-to-leading order.
\section{\bf Procedure}
Consider the $SU(3)$ vector currents,$^{\cite{dgh}}$
\begin{equation}
V^\mu_a = {\bar q}{\lambda_a \over 2}\gamma^\mu q \ \ ,
\label{P1}
\end{equation}
\noindent where $a = 1,\dots ,8$ and $q = (u\ d\ s)$.
For the cases $a=b=3$ and $a=b=8$, the corresponding two-current
time-ordered products can be expressed as$^{\cite{spec}}$
\begin{eqnarray}
\lefteqn{\langle 0|T\left( V^\mu_a (x) V^\nu_b (0)\right)
|0\rangle =} \nonumber \\
& & i \int_0^\infty ds~ \rho_{{\rm V}ab} (s)~( \Box g^{\mu\nu}
- \partial^\mu \partial^\nu) \int {d^4p\over (2\pi)^4}~{e^{-ip\cdot x}\over
p^2 - s + i\epsilon} \ \ ,
\label{P2}
\end{eqnarray}
where the quantities $\rho_{{\rm V}ab} (s)$ are the spin-one vector
spectral functions. The vector current propagators are Fourier
transforms of the spatial two-point functions,
\beq
\Delta^{\mu\nu}_{Vab} (q^2) \equiv i \int d^4 x~e^{iq\cdot x}
\langle 0|T (V^\mu_a (x) V^\nu_b (0)) | 0 \rangle \ \ .
\label{P3}
\eeq

In this paper, we shall use ChPT to calculate the low energy
behavior of the isospin and hypercharge vector-current propagators
$\Delta^{\mu\nu}_{V33} (q^2)$ and $\Delta^{\mu\nu}_{V88} (q^2)$.
Since the underlying current is conserved for each of these,
we have the tensor decomposition
\beq
\Delta^{\mu\nu}_{Vab} (q^2) = (q^\mu q^\nu - q^2 g^{\mu\nu})
\Pi_{ab}(q^2) \qquad (a = b = 3,8) \ \ .
\label{P4}
\eeq
We shall refer to this tensor structure as {\it covariant}.
In the course of calculating $\Delta^{\mu\nu}_{V33} (q^2)$
and $\Delta^{\mu\nu}_{V88} (q^2)$, one encounters a different
class of contributions which are proportional to $g^{\mu\nu}$.
We call these terms {\it noncovariant}.  Of course, the set
of noncovariant terms must ultimately cancel.

\begin{center}
{\bf External Vector Sources}
\end{center}

In accordance with the plan to carry out a general analysis of ChPT to
one-loop, Gasser and Leutwyler employed the so-called background-field
method.$^{\cite{gl1}}$  This had the advantage of generating all possible
${\cal O}(q^4)$ counterterms.  However, since our attention will be strictly
limited to the conserved vector-current propagators $\Pi_{33}$ and
$\Pi_{88}$, we have adopted a rather different approach.  We observe
that it is sufficient to compute all Feynman diagrams of the
type depicted in Fig.~2, {\it i.e.} those with a single
incoming external vector source undergoing a transition to
a single outgoing external vector source.  It will become clear
that this method is both accessible and straightforward to implement.

Let us illustrate the procedure in terms of
the lowest-order chiral lagrangian ${\cal L}^{(2)}$,
\begin{equation}
{\cal L}^{(2)} = {F^2_0\over 4} \Tr \left( D_\mu U D^\mu U^\dagger\right) +
{F^2_0\over 4} \Tr \left(\chi U^\dagger + U\chi ^\dagger\right) \ .
\label{P5}
\end{equation}
In the above, $F_0$ is the pseudoscalar meson decay constant
to lowest order and $\chi = 2B_0 {\bf m}$ is proportional to
the quark mass matrix with
\beq
B_0 = {m_\pi^2 \over 2{\hat m}} = {m_K^2 \over {\hat m}+ m_s } =
{3m_\eta^2 \over 2({\hat m}+ 2m_s)} \ .
\label{P5a}
\eeq
Note that we work in the isospin symmetric limit of $m_u = m_d \equiv
{\hat m}$.  The field variable $U$ is defined in terms of the
pseudoscalar meson fields $\{ \phi_a \}$,
\begin{equation}
U \equiv \exp (i\lambda_a\cdot\phi_a/F_0 ) \ ,\qquad
\label{P6}
\end{equation}
and we construct the covariant derivative $D_\mu U$ in terms of
external vector sources $v_\mu$,
\begin{equation}
D_\mu U \equiv \partial_\mu U + iv_\mu U - iU v_\mu \ \ .
\label{DP}
\end{equation}
The vector source $v_\mu$ has a component $v_\mu^a$ for each of the
$SU(3)$ flavors,
\beq
v_\mu \equiv {\lambda_a \over 2} \cdot v_\mu^a \ \ .
\label{P8}
\eeq

Suppose that we use the above form of ${\cal L}^{(2)}$ to
compute a Feynman diagram of the type in Fig.~2.
For now, consider a term in ${\cal L}^{(2)}$ which is
linear in the vector source $v_\mu$,
\beq
{\cal L}_{\rm int} = - v_\mu^c (x) V^\mu_c (x) \ \ .
\label{P9}
\eeq
The quantity $V^\mu_c (x)$ which is coupled to the external
source $v_\mu^c$ must, by construction, be the corresponding
vector current.  To calculate the given Feynman diagram, we
must work with the second-order operator ${\cal S}_2$,
\beq
{\cal S}_2 -1 = {i^2 \over 2!} \int d^4x ~d^4y \
T\left( {\cal L}_{\rm int}(x) {\cal L}_{\rm int} (y) \right) \ \ .
\label{P10}
\eeq
Upon taking the matrix element for the diagram in Fig.~2
and employing translation invariance, we easily obtain
the result,
\beq
\langle f|{\cal S}_2 -1 |i\rangle = i (2\pi)^4 \delta^{(4)} (q' - q)
{}~\epsilon_\mu^\dagger (q') {\cal M}_{ab}^{\mu\nu} \epsilon_\nu (q) \ ,
\label{P11}
\eeq
where the invariant amplitude ${\cal M}_{ab}^{\mu\nu}$ is found to be
\beq
{\cal M}_{ab}^{\mu\nu} = i \int d^4 x e^{iq\cdot x}
\langle 0 | T( V^\mu_a (x) V^\nu_b (0) | 0 \rangle \equiv
\Delta^{\mu\nu}_{Vab} (q^2) \ .
\label{P12}
\eeq
That is, one just computes the invariant amplitudes for the
class of diagrams in Fig.~2 and the result is the
vector-current propagator of the relevant flavor.  The
only complication is that amplitudes corresponding to
the linear interaction of Eq.~(\ref{P9}) are generally
noncovariant and one must thus include contact terms which
are bilinear in the external vector sources.  For example,
in $SU(3)$ notation the full interaction induced by the
chiral lagrangian ${\cal L}^{(2)}$ is
\beq
{\cal L}_{\rm int} = -f_{abc}\phi_b \partial^\mu \phi_c v_\mu^a
+ {1\over 2} f_{adg} f_{bcg} \phi_c \phi_d v_a^\mu v_b^\nu \ ,
\label{P13}
\eeq
where the two terms are to be calculated respectively in
second and first order perturbation theory.  In the next
section, we shall use this lowest-order interaction to begin
our treatment of the vector-current propagators to one-loop order.

\vspace{0.3cm}

To recap briefly, our calculational program is to use isospin and
hypercharge vector sources $v_3^\mu$ and $v_8^\mu$ as aids in
computing the corresponding vector-current propagators.  We shall
require the chiral lagrangians ${\cal L}^{(2)}$, ${\cal L}^{(4)}$
and ${\cal L}^{(6)}$.  These lagrangians will involve just the
so-called normal vector currents.  In this paper, we will not
consider any of the anomalous currents.  In the
course of computing the one-loop and two-loop Feynman
diagrams, we shall encounter divergent integrals to
which we shall assign precise meaning via
dimensional regularization.  Due to the many terms which
occur, the calculation is both difficult and complicated.
Fortunately, it will be possible to check the correctness of
our results along the way in a variety of ways.
\section{One-loop and Two-loop Analyses}

We shall divide discussion of our calculation into three
separate parts --- loop analysis, counterterm analysis, and
renormalization procedure.  This section begins the process with a
description of one-loop and two-loop amplitudes which are obtained
by using the vector-source approach of Section~2.  The Feynman
diagrams underlying our loop and counterterm analyses are the
ones given in Fig.~(3) and Fig.~(4).  Throughout, we shall denote
all contributions from the chiral ${\cal L}^{(4)}$ by a small
{\it circle} enclosing a cross and those from ${\cal L}^{(6)}$
by a small {\it square} enclosing a cross.  It turns out that
all the Feynman integrals we shall encounter can ultimately be
expressed in terms of just two quantities, which we call $A$ and
${\overline B}_{21}$.  They are defined in Appendix~A, where
we also discuss in detail all such integrals occurring
in our analysis.  In our dimensional regularization treatment of
these quantities, we have been very careful to allow throughout
for sufficiently high powers in the smallness parameter $d - 4$.
As it happens, the only terms in our expansions which
actually contribute are the ones we display in App.~A.

\begin{center}
{\bf One-loop Amplitudes}
\end{center}

The one-loop chiral analysis of the isospin vector-current
propagator was originally carried out by Gasser
and Leutwyler in an $SU(2)$ basis of fields$^{\cite{gl1}}$.
We shall briefly summarize the theoretical analysis with
the $SU(3)$ basis employed in this paper and shall also include
the one-loop expression for the hypercharge propagator.  Our
discussion of the one-loop sector will serve to set the stage for
the two-loop analysis.

At one-loop level, the vector-current propagators are
determined from the Feynman diagrams appearing in Fig.~3.  There
are two amplitudes, each arising from the chiral lagrangian
${\cal L}^{(2)}$, the {\it unitarity} diagram of Fig.~3(a) and the
{\it tadpole} diagram of Fig.~3(b).  For the isospin propagator, one must
sum over contributions from both pion and kaon intermediate states
whereas the hypercharge propagator gets contributions from just kaons.

It is the unitarity diagram which contains the `real physics'
such as the two-particle branch points.  We shall explicitly
consider the hypercharge case as it is somewhat less
complicated.  A straightforward calculation shows the
invariant amplitude for the unitarity diagram of Fig.~3(a) to be
\beqa
\lefteqn{{\cal M}^{\mu\nu}_{88}\bigg|_{\rm unitarity} =
-3iA(m_K^2)g^{\mu\nu}} \nonumber \\
& & - (q^\mu q^\nu - q^2 g^{\mu\nu} ) \left( {iA(m_K^2)
\over 2m_K^2} +
6~i{\overline B}_{21} (q^2, m_K^2) \right)~.
\label{L5}
\eeqa
The unitarity contribution cannot by itself be physical
since it is noncovariant, divergent and scale-dependent.  However,
the tadpole amplitude
\beq
{\cal M}^{\mu\nu}_{88}\bigg|_{\rm tad} = 3iA(m_K^2)g^{\mu\nu}
\label{L6a}
\eeq
is seen to remove the problem of noncovariance in the unitarity amplitude
via cancelation.  The full hypercharge amplitude is then
\beq
\Pi_{88}(q^2)\bigg|_{\rm 1-loop} = -{iA(m_K^2) \over 2m_K^2} -
6~i{\overline B}_{21} (q^2, m_K^2) +
\Pi_{88}^{ct}\bigg|_{\rm 1-loop} \ \ ,
\label{L6b}
\eeq
where $\Pi_{88}^{ct}|_{\rm 1-loop}$ contains the
counterterms generated by the next-order chiral lagrangian
${\cal L}^{(4)}$.$^{\cite{gl2}}$  This latter contribution is
shown in Fig.~3(c) and will be considered in Section~4.

The above discussion has concerned determination of just the hypercharge
vector-current propagator at one-loop order.  For completeness, we
display without further discussion the one-loop amplitude for the
isospin case ($a = b = 3$),
\beqa
\lefteqn{\Pi_{33} (q^2)\bigg|_{\rm 1-loop} =
-{iA(m_\pi^2) \over 3m_\pi^2} - {iA(m_K^2) \over 6m_K^2}} \nonumber \\
& & - 4~i{\overline B}_{21} (q^2, m_\pi^2) - 2~i{\overline B}_{21}
(q^2, m_K^2) + \Pi_{33}^{ct}\bigg|_{\rm 1-loop} \ \ .
\label{L7}
\eeqa
A simple check of the calculation is obtained by passing to
the $SU(3)$ limit of $m_\pi = m_K$.   As expected, the isospin
and hypercharge amplitudes agree.

\begin{center}
{\bf Two-loop Amplitudes}
\end{center}

In some sense, the content of this subsection must parallel that
just given.  One simply calculates the set of amplitudes
associated with Feynman diagrams at two-loop order.  However, in
perturbative quantum field theory, things get worse
as one proceeds to higher orders.  Such is the case here.
Instead of $2$ types of non-counterterm amplitudes as in Fig.~3(a)-(b),
there are now $9$ non-counterterm amplitudes shown in
Fig.~4(a)-(i).  Of these, all but the `eyeglass' diagram
of Fig.~4(i) may be considered as corrections of one kind or another
to the one-loop diagrams of Fig.~3.  A detailed account describing
evaluation of these diagrams is clearly too lengthy a process
to reproduce here.  However, we can perhaps convey to the reader
some useful insights with the following observations:
\begin{enumerate}
\item Since the different $\pi$, $K$ and $\eta$ flavors can
generally occur in each loop, there are many terms contributing
to both the isospin and hypercharge two-loop amplitudes.
\item A number of diagrams have noncovariant ($g_{\mu\nu}$)
contributions but they all turn out to cancel, as they must.
\item Interestingly, some diagrams contain terms in which a divergent
quantity is multiplied by a nonpolynomial function.  Such terms
are potentially disasterous to the calculational program since they
cannot be removed by the usual renormalization procedure.
However, they cancel totally.
\item Two-loop amplitudes contain a number of contributions from
the chiral lagrangian ${\cal L}^{(4)}$, in addition to those from
${\cal L}^{(2)}$.
\end{enumerate}

The final comment motivates introduction at this point of
${\cal L}^{(4)}$, the next-to-lowest order chiral
lagrangian.$^{\cite{gl2}}$ In principle, the entire ${\cal O}(q^4 )$
lagrangian has three parts.  The first has terms containing the
meson field variable $U$,
\begin{eqnarray}
{\cal L}^{(4)} &=& \sum_{i=1}^{10}~L_i O_i \nonumber \\
&=& L_1 \left[ \Tr \left( D_\mu U D^\mu U^\dagger\right)\right]^2
+ L_2 \Tr \left( D_\mu U D_\nu U^\dagger \right)\cdot
\Tr \left( D^\mu U D^\nu U^\dagger\right) \nonumber \\
&+& L_3 \Tr \left( D_\mu U D^\mu U^\dagger D_\nu U D^\nu U^\dagger\right)
\nonumber \\
&+& L_4 \Tr \left( D_\mu U D^\mu U^\dagger\right) \Tr \left( \chi
U^\dagger+ U \chi^\dagger\right) \label{L8} \\
&+& L_5 \Tr \left( D_\mu U D^\mu U^\dagger\left(\chi U^\dagger + U
\chi^\dagger\right)\right)
 + L_6 \left[ \Tr \left( \chi U^\dagger + U \chi^\dagger\right)\right]^2
\nonumber \\
&+& L_7 \left[ \Tr\left( \chi^\dagger U - U\chi^\dagger\right)\right]^2
+ L_8 \Tr \left( \chi U^\dagger \chi U^\dagger + U\chi^\dagger
U\chi^\dagger\right) \nonumber \\
&+& i L_9 \Tr \left( L_{\mu\nu} D^\mu U D^\nu U^\dagger + R_{\mu\nu} D^\mu
U^\dagger D^\nu U\right) \nonumber \\
&+& L_{10} \Tr \left( L_{\mu\nu} U R^{\mu\nu} U^\dagger\right). \nonumber
\end{eqnarray}
The second has additional contributions occurring at the same
order but which contain only external sources,
\begin{equation}
{\cal L}_{\rm ext} = H_1\Tr \left( L_{\mu\nu} L^{\mu\nu} +
R_{\mu\nu}R^{\mu\nu} \right) + H_2\Tr\left(\chi^\dagger \chi\right) \ \ .
\label{L9}
\end{equation}
In the above, $L_{\mu\nu},R_{\mu\nu}$ are the field-strength tensors
\begin{eqnarray}
L_{\mu\nu} &=& \partial_\mu \ell_\nu - \partial_\nu \ell_\mu
+ i [\ell_\mu , \ell_\nu ] \nonumber \\
R_{\mu\nu} &=& \partial_\mu r_\nu - \partial_\nu r_\mu
+ i [ r_\mu , r_\nu ] \ \ .
\label{L10}
\end{eqnarray}
The third consists of two terms which vanish upon using the equations
of motion,
\begin{equation}
{\cal L}_{\rm eq-motn}=G_1 \Tr\left({\cal E}\cdot{\cal E}\right) + G_2
\Tr\left({\cal E}\cdot i\left(\chi^\dagger U-U^\dagger\chi\right)\right)\ ,
\end{equation}
where
\begin{equation}
{\cal E} \equiv i\left( 2 D_\mu\left(U^\dagger D^\mu U\right)
+\chi^\dagger U-U^\dagger\chi
-{1\over 3} \Tr \left(\chi^\dagger U-U^\dagger\chi\right) 1 \right)\ .
\end{equation}
In applications up to one-loop order, these terms can be neglected since
they contribute only through tree diagrams where the equations of motion
${\cal E}=0$ may be applied. This is no longer true for a two-loop calculation.
The terms in ${\cal L}_{\rm eq-motn}$ may in general contribute a polynomial to
the amplitude, which in turn could be absorbed into the coupling constants
of the ${\cal O}(q^6)$ counterterm lagrangian.$^{\cite{e93}}$ In our
application
we have kept these terms at all stages of the calculation. Although present in
intermediate steps, however, the effect of $G_1$, $G_2$ cancels
completely in the final result.

For the hypercharge vector-current propagator the result
of working out all the two-loop Feynman diagrams is
\beqa
\lefteqn{\Pi_{88} (q^2)\bigg|_{\rm 2-lp} = {1\over F_0^2} \left[
- 24 q^2 L_9 i{\overline B}_{21} (q^2, m_K^2)
+ 2 iA(m_K^2) \left[ 6L_{10} + (6 - {q^2 \over m_K^2})L_9
\right]\right. } \nonumber \\
& & \left. + q^2 \left[ {1\over 8}\left({iA(m_K^2)\over m_K^2}\right)^2
+ 3{iA(m_K^2)\over m_K^2}i{\overline B}_{21}(q^2 , m_K^2)
+ 18 (i{\overline B}_{21}(q^2 , m_K^2))^2 \right] \right. \nonumber \\
& & \left. - 3{m_K^2\over q^2} i{\overline B} (q^2, m_K^2)
\left[ 8 (m_\pi^2 + 2 m_K^2 )(L_4 - 2 L_6 ) + 8 m_K^2 (L_5 - 2L_8 )
\vphantom{{A\over m_K^2}}\right. \right. \nonumber \\
& &\left. \left. - { 1 \over 3}iA(m_\eta^2) \right] \right] \
+\  \Pi_{88}^{ct}\bigg|_{\rm 2-loop} \ \ ,
\label{L11}
\eeqa
where we display only implicitly the two-loop counterterm
contribution $\Pi_{88}^{ct}|_{\rm 2-loop}$ of Fig.~4(j)
and defer its calculation to Sect.~4.  Observe how the above
result reinforces the statement made earlier
that all Feynman integrals in our analysis are reducible to
just the two quantities $A$ and ${\overline B}_{21}$.

To summarize, up to this point we have determined the isospin
and hypercharge amplitudes through two-loop order,$^{\cite{hlm}}$
\beq
\Pi_{ab} (q^2) = \Pi_{ab} (q^2)\bigg|_{\rm 1-loop}
+ \Pi_{ab} (q^2)\bigg|_{\rm 2-loop} \qquad (a = b = 3,8) \ .
\label{L12}
\eeq
These results cannot yet be accepted as physically
meaningful due to the presence of divergent terms and to the
explicit dependence on an arbitrary scale $\mu$ which is an
artifact of splitting loop and counterterm contributions.

\section{Enumeration of Counterterms}

The difficulties of divergence and scale-dependence in the
1-loop and 2-loop amplitudes are overcome by making appropriate
use of counterterms.  It is necessary to first generate a list at both
1-loop and 2-loop order of all counterterms which can possibly contribute.
Carrying out this process for the ${\cal O}(q^6)$ counterterms
will be the primary objective of this section.

Let us begin by reminding ourselves of which counterterms appear at
${\cal O}(q^4)$ order.$^{\cite{gl2}}$  Out of a total of $12$ possible
operators in the chiral lagrangians ${\cal L}^{(4)}$ of Eq.~(\ref{L8})
and ${\cal L}_{\rm ext}$ of Eq.~(\ref{L9}), only the $L_{10}$ and
$H_1$ terms turn out to have nonzero matrix elements taken between
single external vector-source lines.  One easily finds the
${\cal O}(q^4)$ counterterm amplitude to have the flavor-independent
and covariant form,
\beq
\Pi^{ct}_{ab}\bigg|_{1-loop} = -2(L_{10} + 2H_1)
\qquad \qquad (a = b = 3,8) \ \ .
\label{C1}
\eeq
As noted earlier, the ${\cal O}(q^4)$ counterterm contribution appears
in Fig.~3(c).

\begin{center}
{\bf Determination of the ${\cal O}(q^6)$ Counterterms}
\end{center}

Since the construction to follow will be somewhat formal in nature,
it is important to keep in mind certain tenets of ChPT for guidance.
Thus --- we have found the two-loop calculations of $\Pi_{33}$ and $\Pi_{88}$
to produce ultraviolet divergences. The low energy counting scheme of ChPT
implies that these divergencies can be absorbed into the coupling constants
of counterterms of order $q^6$, thus renormalizing the
amplitude.$^{\cite{Wei79}}$  However, more generally speaking all possible
counterterms of order $q^6$ have to be included in order to obtain the
most general amplitude consistent with chiral symmetry.  This is irrespective
of whether divergencies occur in the loop amplitude or not.

In addition to those counterterms that contribute to $\Pi_{33}$
and $\Pi_{88}$, we shall also construct the set of ${\cal O}(q^6)$
counterterms which enter the two-loop analysis of the process
$\gamma\gamma\rightarrow\pi^0\pi^0$ and $\eta\rightarrow\pi^0\gamma\gamma$.
Our motivation stems in part from a recent analysis at two-loop order in
ChPT of the first of these reactions.$^{\cite{BGS94}}$  In that work,
however, no reference is made to a complete basis of ${\cal O}(q^6)$
counterterms.  Our aim here will be to establish the precise
relation between ${\cal O}(q^6)$ counterterms which contribute to both
the $\gamma\gamma\rightarrow \pi^0\pi^0$ amplitudes and the vector current
two-point function.\footnote{Throughout this article we work in chiral
SU(3). Comparison to the constants appearing in Ref.~\cite{BGS94} is
therefore only possible after the SU(2) limit has been taken.}
In addition, we hope that the details given below
will serve to simplify comparison with future works in this subject.

Our first step is to derive a list of counterterms contributing
{\it a priori} to the processes under consideration. Since we consider only
couplings to neutral pseudoscalar mesons, the analysis simplifies
considerably. There are no one-particle reducible diagrams involving
${\cal O}(q^6)$ counterterms. The reason is simply that to lowest order,
$q^2$, the vector source couples to an even number of Goldstone bosons.
We may therefore switch off the external fields except for the vector
sources $v_\mu^a$.  Furthermore, since all flavour matrices involved are
diagonal, the situation can be described by
\begin{eqnarray}
a_\mu &=&0~,\qquad p=0~,\qquad s= {\bf m}\ , \nonumber\\
\phi  &=&{\rm diagonal}\quad \rightarrow\quad U={\rm diagonal}\ \ .
\label{simplify}
\end{eqnarray}
To proceed, we choose a set of building blocks transforming homogeneously under
chiral transformations. The invariants can then be constructed easily by taking
traces in flavor space.  Our construction will closely follow the
work in Refs.~\cite{EGPR89,EKW93}. The building blocks we use are
\begin{eqnarray}
\chi_\pm &=& u (U^\dagger\chi\pm\chi^\dagger U) u^\dagger \ ,\nonumber\\
u_\mu    &=& i u^\dagger D_\mu U u^\dagger \ ,\nonumber\\
w_{\mu\nu}&=& \nabla_\mu u_\nu+\nabla_\nu u_\mu \ ,\label{buildbl}\\
f_{\pm \mu\nu}&=&u R_{\mu\nu} u^\dagger \pm u^\dagger L_{\mu\nu} u \ ,
\nonumber
\end{eqnarray}
as well as the totally antisymmetric tensor $\epsilon_{\mu\nu\rho\sigma}$.
All the quantities $U$, $\chi$, $D_\mu U$ and the field strengths
$L_{\mu\nu}$, $R_{\mu\nu}$ have been defined in previous sections
and $u$ is the `square-root' of the matrix-field $U$,
\begin{equation}
U=u\cdot u \ \ .
\end{equation}
Under a chiral $SU(3)$ $\otimes$ $SU(3)$ transformation, the matrix
$u$ transforms as
\begin{equation}
u(\phi)\rightarrow g_R u(\phi) h^\dagger(\phi)=h(\phi) u(\phi)
g_L^\dagger \ \ ,
\label{mon}
\end{equation}
where the nonlinear transformation $h(\phi)$ which first appears in
Eq.~(\ref{mon}) is, in fact, defined by this same
equation.$^{\cite{CWZ69,CCWZ69}}$
As a consequence, the left-chiral current $u_\mu$ transforms homogeneously,
\begin{equation}
 u_\mu\rightarrow h(\phi) u_\mu(\phi) h^\dagger(\phi)\ \ .
\label{tue}
\end{equation}
Given some object ${\cal O}$, a covariant derivative $\nabla_\mu {\cal O}$
which transforms in like manner to $u_\mu$ can be defined as
\begin{equation}
\nabla_\mu {\cal O}\equiv \partial_\mu {\cal O} +[\Gamma_\mu,{\cal O}]\ \ ,
\end{equation}
where
\begin{equation}
\Gamma_\mu={1\over 2} \left( u^\dagger[\partial_\mu-i(v_\mu+a_\mu)]u+
       u[\partial_\mu-i(v_\mu-a_\mu)]u^\dagger\right) \ \ .
\end{equation}
Here $v_\mu$, $a_\mu$ are the matrix-valued external fields with spin-one
introduced in Sect. 2.

The building blocks of Eq.~(\ref{buildbl}) as well as covariant derivatives
thereof thus transform according to
\begin{equation}
{\cal O}\rightarrow h(\phi) {\cal O} h^\dagger(\phi)\ \ .
\end{equation}
The most general set of counterterms of order $q^6$ is then constructed by
multiplying together the building blocks in Eq.~(\ref{buildbl}) up to
the appropriate order in the low energy expansion and then performing
flavor traces like
\begin{equation}
\Tr( {\cal O}_i{\cal O}_j \ldots),\quad \ldots,\quad
\Tr( {\cal O}_i{\cal O}_j \ldots) \Tr({\cal O}_k \ldots)\ldots
\Tr( {\cal O}_l \ldots)\ \ .
\label{invar}
\end{equation}
In Eq.~(\ref{invar}), Lorentz indices are to be contracted and symmetrizations
performed such that the discrete symmetries $P$ and $C$ are conserved. The
transformation properties of the various building blocks under $P$ and $C$
can be found in Ref.~\cite{EGPR89}. We also note that all quantities in
Eq.~(\ref{buildbl}), except for $\chi_\pm$, are traceless.

The procedure just described for constructing the invariants of order $q^6$
leads to a set of terms which is in general redundant. In principle this
set of terms can be reduced further by the use of the equations of motion,
trace and epsilon identities$^{\cite{FS94}}$ as well as by partial
integrations. We shall not elaborate further on such procedures, since
the case we are interested in is much simpler and does not require
that level of analysis.

Let us return to the restricted situation as
characterized in Eq.~(\ref{simplify}). Under these circumstances
the building blocks of Eq.~(\ref{buildbl}) are transformed into
\beq
\begin{tabular}{lcl}
$\chi_\pm$ &=& $2 B_0 {\bf m}(U^\dagger\pm U)$ , \\
$u_\mu$ &=& $i u^\dagger \partial_\mu u^\dagger$ , \\
$w_{\mu\nu}$ &=& $\partial_\mu u_\nu+\partial_\nu u_\mu$ , \\
\end{tabular}
\qquad
\begin{tabular}{lcl}
$f_{+ \mu\nu}$ &=& $2 (\partial_\mu v_\nu-\partial_\nu v_\mu)$ \ ,\\
$f_{- \mu\nu}$ &=& $0$\ \ .\\
\phantom{}  & & \phantom{}   \\
\end{tabular}
\label{simp}
\eeq
The only nontrivial objects at our disposal are therefore $\chi_+$,
$f_{+ \mu\nu}$ and covariant derivatives thereof as well as the totally
antisymmetric $\epsilon$-tensor. The covariant derivative is also simplified,
and is explicitly given as
\begin{equation}
\nabla_\mu {\cal O}=\partial_\mu {\cal O} -i[v_\mu, {\cal O}]\ \ .
\end{equation}
Since only the diagonal isospin and hypercharge components $a=3,8$ are
studied in this work, it follows that $\nabla_\mu \chi_+ \sim \nabla_\mu M=0$.
In the same manner, we see that $f_{+ \mu\nu}$ must be linear in the
vector source $v_\mu$, so two powers of $f_{+ \mu\nu}$ are needed.
As regards the operators $\chi_-$ and $w_{\mu\nu}$, we note that $\chi_-$
involves only odd numbers of mesons and that $w_{\mu\nu}$is parity-odd and
hence cannot lead to an operator of the form we are looking for. Finally,
the remaining two powers of `small' momenta as well as the two
pseudoscalar meson fields can only be provided by two powers of $u_\mu$
or one factor of $\chi_+$.
We are thus left with an expression of the form
\begin{eqnarray}
{\cal L}^{(6)} &=& {1\over F_0^2}\left(
K_1 \Tr(\nabla_\lambda f_+^{\mu\nu} \nabla^\lambda f_{+ \mu\nu})
+K_2 \Tr(\nabla_\mu f_+^{\mu\nu} \nabla^\lambda f_{+ \lambda\nu})
\right. \nonumber \\
& & \left. \phantom{x}+K_3 \Tr(f_{+ \mu\nu}f_+^{\mu\nu} \chi_+)
+K_4 \Tr(f_{+ \mu\nu}f_+^{\mu\nu})\Tr(\chi_+) \right. \nonumber \\
& & \left. \phantom{x}
+ K_{10} \Tr(f_{+ \mu\nu}f_+^{\mu\nu} u^\rho u_\rho)
+ K_{11} \Tr(f_{+ \mu\nu}f_+^{\mu\rho} u^\nu u_\rho) + \ldots
\vphantom{1\over F_0^2}\right)  \ .
\label{L6}
\end{eqnarray}
The ellipses in this equation denote all possible independent operators
which can arise from the terms containing $K_{10}$ and $K_{11}$ by changing
the order of the building blocks $f_{+ \mu\nu}$ and $u_\rho$ within the
flavour trace and/or by taking two flavour traces.  We shall refrain from
writing down such contributions explicitly since they involve only diagonal
flavour traces and thus lead to interactions of the type
\begin{eqnarray}
{\cal L}_{\rm der}^{(6)'} &=& {4 e^2\over F_0^4} F_{\mu\nu}F^{\mu\rho}
\left[ (K_{10}' g_{\sigma\tau} g^{\nu\rho}+K_{11}'g_\sigma^\nu g_\tau^\rho)
\Tr(Q^2 \partial^\sigma \phi \partial^\tau \phi) \right. \nonumber \\
& &\left. + (K_{12}' g_{\sigma\tau} g^{\nu\rho}+K_{13}'g_\sigma^\nu
g_\tau^\rho)
\Tr(Q^2)\Tr(\partial^\sigma \phi \partial^\tau \phi) \right. \label{jk1} \\
& &\left. +(K_{14}' g_{\sigma\tau} g^{\nu\rho}+K_{15}'g_\sigma^\nu g_\tau^\rho)
\Tr(Q \partial^\sigma \phi) \Tr(Q \partial^\tau \phi)
\vphantom{\left(K_{10}' g_{\sigma\tau} g^{\nu\rho}\right)}\right] \ . \nonumber
\label{L6der}
\end{eqnarray}
In the above, we define the electromagnetic field strength
\beq
F_{\mu\nu} \equiv \partial_\mu A_\nu - \partial_\nu A_\mu \ \ ,
\label{jk2}
\eeq
where $A_\mu$ is the photon field, and take
\begin{equation}
v_\mu=e Q A_\mu \ , \qquad Q={1\over3} {\rm diag}(2,-1,-1) \ \ .
\label{jk3}
\end{equation}

It is now straightforward to determine the counterterm contributions
to $\Pi_{33}$ and $\Pi_{88}$ which arise from the lagrangian
${\cal L}^{(6)}$.  We obtain
\beq
\Pi^{ct}_{33}(q^2)\bigg|_{\rm 2-loop} = -{4\over F_0^2} \left[ q^2
(2 K_1+ K_2)
+4 m_\pi^2 K_3 + 4 (2m_K^2+m_\pi^2) K_4 \right]
\label{final1}
\eeq
and
\beq
\Pi^{ct}_{88} (q^2)\bigg|_{\rm 2-loop} = -{4\over F_0^2} \left[
q^2 (2 K_1+ K_2)
+ 4 m_\eta^2 K_3 + 4 (2m_K^2+m_\pi^2) K_4 \right] \ ,
\label{final2}
\eeq
where the lowest order expressions for squares of the meson masses
({\it cf.} Eq.~(\ref{P5a})) have been employed. Note that the contribution
of $q^6$ counterterms to the difference, $\Pi_{33}-\Pi_{88}$, is a constant
depending only on $K_3$. Also, from the four terms $K_1$, ... $K_4$ which
may contribute a priori, only three independent combinations actually
appear. The renormalization of coupling constants
such as $K_1$, ... $K_4$ as well as their properties following from the
renormalization group equations will be discussed in Sect.~5.

As for the two-meson two-photon processes, we first note a relation
between $\gamma\gamma\rightarrow\pi^0\pi^0$ and
$\eta\rightarrow\pi^0\gamma\gamma$.
Upon evaluating the flavour traces in Eqs. (\ref{L6}),(\ref{L6der})
we easily
see that as long as the pieces which violate the Zweig rule are
neglected ({\it i.e.} those terms with two flavour traces), then
the amplitudes for $\gamma\gamma\rightarrow \pi^0\pi^0$ and
$\eta \rightarrow \pi^0 \gamma\gamma$ will depend on the {\it same}
linear combination of the $\{K_i\}$.$^{\cite{Kne94}}$  Let us not pursue
this line of reasoning any further, but instead concentrate in
the following on the $\gamma\gamma\rightarrow \pi^0\pi^0$ reaction.
The $\gamma\gamma\rightarrow \pi^0\pi^0$ matrix element can be written as
\begin{equation}
_{\rm out}\langle \pi^0(p_1)\pi^0(p_2) |\gamma(q_1,\epsilon_1)
\gamma(q_2,\epsilon_2)\rangle_{\rm in} = i (2\pi)^4 \delta^{(4)}(P_f-P_i)
e^2 \epsilon_1^\mu \epsilon_2^\nu V_{\mu\nu}
\label{jk5}
\end{equation}
with
\begin{equation}
V^{\mu\nu}=A(s,t,u) T_1^{\mu\nu} +B(s,t,u) T_2^{\mu\nu} \ \ .
\label{jk6}
\end{equation}
The tensors $T_{1,2}^{\mu\nu}$ are defined in Ref.~\cite{BGS94}, and we
shall adopt the notation and conventions used there.  It is
straightforward to determine the contribution of ${\cal O}(q^6)$
counterterms to the Lorentz invariant amplitudes $A$, $B$, and we find
\begin{eqnarray}
& & A^{\rm ct}  (s,t,u) = {4 \over F_0^4} \left[ -8 M_\pi^2
\Bigl(-{5\over 9} K_3-{4\over 3} K_4+{5\over 9} K_{10}'+{4\over 3} K_{12}'
+K_{14}'\Bigr) \right. \nonumber \\
& &\left. + ~ s \Bigl( {5\over 18} (8 K_{10}' + K_{11}')
+ {2\over 3} (8K_{12}' + K_{13}') + 4K_{14}'+ {1\over 2} K_{15}' \Bigr)
\right] \ ,\nonumber\\
& & B^{\rm ct}(s,t,u) = -{1\over F_0^4} \left[ {5\over 9} K_{11}'+{4\over 3}
K_{13}'+K_{15}' \right] \ . \label{jk7}
\end{eqnarray}
This completes our discussion of the counterterm contribution to
the $\gamma\gamma\rightarrow \pi^0\pi^0$ reaction.
Observe that the derivative lagrangian also contributes a
constant piece to the amplitude $A$. The information contained in the
two-loop expression for the vector propagator is thus not sufficient
to fix the constant shift in A at this order of the low energy expansion.

\section{Renormalization Prescription}

Now that we have a complete set of ${\cal O}(q^4)$
and ${\cal O}(q^6)$ counterterms, the next step is
to formulate a procedure which allows the divergences
to be subtracted and the scale-dependence removed from
the calculation.

\begin{center}
{\bf Subtraction Procedure}
\end{center}

There exist different ways of implementing
these steps, {\it e.g.} minimal subtraction, {\it etc}.
We have found it most convenient to subtract in the quantity
${\overline\lambda}$.  This amounts to expanding the
various counterterms in decreasing powers of ${\overline\lambda}$.
In the following, let $K$ denote any of the ${\cal O}(q^6)$
counterterms and $L$ represent any of the ${\cal O}(q^4)$
counterterms.  For the former, we write
\beqa
K &=& \mu^{2(d - 4)} \sum_{n=2}^{-\infty} \ K^{(n)}(\mu)
{}~{\overline\lambda}^n \nonumber \\
&=& \mu^{2(d - 4)} \left[ K^{(2)}(\mu) ~{\overline\lambda}^2
+ K^{(1)}(\mu) ~{\overline\lambda} + K^{(0)}(\mu) + \dots \right] \ \ ,
\label{R1a}
\eeqa
whereas the latter is given by
\beqa
L &=& \mu^{d - 4} \sum_{n=1}^{-\infty} \ L^{(n)}(\mu)
{}~{\overline\lambda}^n \nonumber \\
&=& \mu^{d - 4} \left[ L^{(1)}(\mu) ~{\overline\lambda} +
L^{(0)}(\mu) + L^{(-1)}(\mu) ~{\overline\lambda}^{-1} + \dots \right] \ \ .
\label{R1b}
\eeqa
In our two-loop analysis, we shall actually need only the $K^{(2)}$,
$K^{(1)}$, $K^{(0)}$ and the $L^{(1)}$, $L^{(0)}$, $L^{(-1)}$
counterterms.

The definition of ${\cal O}(q^4)$ counterterms used here has
a simple association with that originally introduced by Gasser
and Leutwyler in their papers on one-loop chiral perturbation
theory.  We restrict our attention to Ref.~\cite{gl2} since,
like this paper, it too employs an $SU(3)$ basis.  Then starting from
Eq.~(7.25) of Ref.~\cite{gl2},
\beqa
L_i^{({\rm G-L})} &=& \Gamma_i \lambda + L_i^r \qquad
(i = 1,\dots,10)\nonumber \\
H_i^{({\rm G-L})} &=& \Delta_i \lambda + H_i^r \qquad (i = 1,2) \ ,
\label{R1c}
\eeqa
we see that our $\{ L_i^{(n)}\}$ and $\{ H_i^{(n)}\}$ are related to
the Gasser-Leutwyler parameters by
\beqa
L_i^{(1)} &=& \Gamma_i \qquad {\rm and} \qquad L_i^{(0)} = L_i^r
\qquad (i = 1, \dots ,10) \nonumber \\
H_i^{(1)}  &=& \Delta_i \qquad {\rm and} \qquad H_i^{(0)}  = H_i^r
\qquad (i = 1,2)\ \ .
\label{R1d}
\eeqa

\begin{center}
{\bf Scale Dependence}
\end{center}

Having chosen a subtraction procedure, we must next determine
the scale dependence of the counterterm coefficients.  This
is done by first noting that although the full counterterm is
independent of scale,
\beq
\mu{dK \over d\mu} = 0 \qquad {\rm and} \qquad
\mu{dL \over d\mu} = 0 \ \ ,
\label{R2}
\eeq
the $\{ K^{(n)}\}$ and $\{ L^{(n)}\}$ must themselves
generally be scale-dependent in order to compensate the scale-dependence of
the prefactors, $\mu^{2(d - 4)}$ in Eq.~(\ref{R1a}) and
$\mu^{d - 4}$ in Eq.~(\ref{R1b}).  Insertion
of Eq.~(\ref{R1a}) into the first equation of Eq.~(\ref{R2}) yields
an infinite sequence of differential relations for the ${\cal O}(q^6)$
counterterms,
\beqa
\mu{dK^{(2)} \over d\mu} &=& 0 \nonumber \\
\mu{dK^{(1)} \over d\mu} &=& -{2\over 16\pi^2}K^{(2)} \label{R3} \\
\mu{dK^{(0)} \over d\mu} &=& {2\over 16\pi^2}
\left[ {C\over 16\pi^2}K^{(2)} - K^{(1)} \right]  \ \ .\nonumber \\
& & \vdots \nonumber
\eeqa
where, in terms of the Euler constant $\gamma$, we define
\beq
C \equiv  {1\over 2} \left( \ln (4\pi) - \gamma + 1 \right)\ \ .
\label{R4}
\eeq
An analogous but distinct set of relations is obtained for the
${\cal O}(q^4)$ counterterms by inserting Eq.~(\ref{R1b}) into the
second equation of Eq.~(\ref{R2}),
\beqa
\mu{dL^{(1)} \over d\mu} &=& 0 \nonumber \\
\mu{dL^{(0)} \over d\mu} &=& -{1\over 16\pi^2}L^{(1)} \label{R5} \\
\mu{dL^{(-1)} \over d\mu} &=& {1\over 16\pi^2}
\left[ {C\over 16\pi^2}L^{(1)} - L^{(0)} \right]  \ \ .\nonumber \\
& & \vdots \nonumber
\eeqa

It is straightforward to sequentially solve relations of the
type given above, and we find for the first few of the $\{ K^{(n)}\}$,
\beqa
K^{(2)}(\mu) &=& K^{(2)}(\mu_0) \qquad ({\rm independent\ of}\ \mu) \nonumber
\\
K^{(1)}(\mu) &=& K^{(1)}(\mu_0) + {K^{(2)}\over 16\pi^2}
\ln\left( {\mu_0^2 \over \mu^2}\right) \nonumber \\
K^{(0)}(\mu) &=& K^{(0)}(\mu_0) + {1\over 16\pi^2} \left[
\left( K^{(1)}(\mu_0) - {C K^{(2)} \over 16\pi^2} \right)
\ln\left( {\mu_0^2 \over \mu^2}\right) \right. \nonumber \\
& & \left. + {K^{(2)} \over 32\pi^2} \ln^2\left( {\mu_0^2 \over
\mu^2}\right) \right] \ \ ,
\label{R6}
\eeqa
whereas the sequence of $\{ L^{(n)}\}$ begins as
\beqa
L^{(1)}(\mu) &=& L^{(1)}(\mu_0) \qquad ({\rm independent\ of}\ \mu) \nonumber
\\
L^{(0)}(\mu) &=& L^{(0)}(\mu_0) + {L^{(1)}\over 32\pi^2}
\ln\left( {\mu_0^2 \over \mu^2}\right) \nonumber \\
L^{(-1)}(\mu) &=& L^{(-1)}(\mu_0) + {1\over 32\pi^2} \left[
\left( L^{(0)}(\mu_0) - {C L^{(1)} \over 16\pi^2} \right)
\ln\left( {\mu_0^2 \over \mu^2}\right) \right. \nonumber \\
& & \left. + {L^{(1)} \over 64\pi^2} \ln^2\left( {\mu_0^2 \over
\mu^2}\right) \right] \ \ .
\label{R7}
\eeqa

Up to this point, the $\{ K^{(n)}\}$ and $\{ L^{(n)}\}$ are
arbitrary.  In Sect.~6, we shall show how to constrain the
counterterms which are relevant to our calculation by imposing
conditions of finiteness and scale-independence.

\begin{center}
{\bf Effect of Mass Renormalization}
\end{center}

For the final topic in this Section, we consider renormalization
of the meson mass parameters.  This intermediate step
turns out to considerably simplify the form of both the
isospin and hypercharge vector-current propagators
$\Pi_{33}$ and $\Pi_{88}$.  For definiteness, we shall again
explicitly study just $\Pi_{88}$ and thus restrict our attention
to the renormalization of the kaon mass parameter.

Recall in Ref.~\cite{gl2} that for kaons,
the relation between the bare mass parameter $m_K$ and
the radiatively-corrected quantity $M_K$ was shown to be
\beq
m_K^2 = M_K^2 + \Delta_K \ \ ,
\label{R8}
\eeq
where
\beqa
\Delta_K &=& {1\over F_0^2} \left[ 8M_K^2 (M_\pi^2 + 2 M_K^2 )
(L_4^{(0)} - 2 L_6^{(0)} ) \right. \nonumber \\
& & \left. + 8 M_K^4 (L_5^{(0)}  - 2L_8^{(0)}  ) -
{M_K^2 M_\eta^2 \over 48\pi^2} \ln \left( {M_\eta^2 \over \mu^2}\right)
\right] \ .
\label{R9}
\eeqa
For the two-loop amplitude, one can replace $m_K^2$ by $M_K^2$
everywhere since the error made occurs in still higher order.
However, at one-loop order Eqs.(\ref{R8}),(\ref{R9}) must be
utilized everywhere, {\it i.e.} both where $m_K^2$ appears explicitly
and also where it occurs implicitly within a function.  The effect
of this substitution is
\beq
\Pi_{88} (q^2 , m_K^2)\bigg|_{\rm 1-loop} =
\Pi_{88} (q^2 , M_K^2)\bigg|_{\rm 1-loop}
+ {3\Delta_K \over q^2} i{\overline B}(q^2 , M_K^2) \ .
\label{R10}
\eeq
It can be shown that the additive term in Eq.~(\ref{R10}) has
the effect of cancelling the final term in the form for
$\Pi_{88} (q^2)|_{\rm 2-loop}$ appearing in Eq.~(\ref{L11}).
This leaves us with the more compact expression,
\beqa
\lefteqn{\Pi_{88} (q^2, M_K^2 )\bigg|_{\rm 2-loop} =
{1\over F_0^2} \left[ - 24L_9 q^2 i{\overline B}_{21} (q^2, M_K^2)
\right.} \nonumber \\
& & \left. + 2iA(M_K^2) \left[ 6L_{10} + (6 - {q^2 \over M_K^2})L_9
\right] + q^2 \left[ {1\over 8}\left({iA(M_K^2)\over M_K^2}
\right)^2 \right. \right. \nonumber \\
& & \left. \left. + 3{iA(M_K^2)\over M_K^2}i{\overline B}_{21}(q^2 , M_K^2)
+ 18 (i{\overline B}_{21}(q^2 , M_K^2))^2 \right] \right] .
\label{R11}
\eeqa
The isospin amplitude behaves analogously, except now both
pion and kaon mass renormalizations must be administered.

\section{Results}

We now have sufficient theoretical machinery to obtain
our final representations for the isospin and hypercharge
vector-current propagators.  The first step in this process
is to expand the one-loop and two-loop expressions obtained
earlier in powers of the singular quantity ${\overline \lambda}$.
When combined with the counterterm contributions which were derived
in the previous section, the full amplitude is rendered finite
and free of the arbitrary scale $\mu$ by appropriate choices
of the renormalization constants contained within the counterterms.
The discussion becomes clearest if we first consider the one-loop case in
some detail, and then touch on the high points of the more
complicated two-loop sector.

\begin{center}
{\bf Renormalization at One-loop Order}
\end{center}

It is again convenient to refer to the simpler hypercharge
one-loop amplitude to illustrate the process.  From
Eq.~(\ref{L6b}) of Sect.~3 and Eq.~(\ref{a3}) of App.~A, we have
\beqa
\Pi_{88}(q^2)\bigg|_{\rm 1-loop} &=&
- 2L_{10} - 4H_1 -{iA(M_K^2) \over 2M_K^2} -
6~i{\overline B}_{21} (q^2, M_K^2) \nonumber \\
&=&  -\mu^{d - 4} \left[ \left( 1 + 2L_{10}^{(1)} + 4 H_1^{(1)} \right)
\cdot {\overline \lambda} \right. \nonumber \\
& & \left. + 2L_{10}^{(0)}(M_K^2) + 4 H_1^{(0)}(M_K^2)
\vphantom{\ln {M_K^2 \over \mu^2}} \right]
- 6~i{\overline B}_{21} (q^2, M_K^2) \ \ .
\label{F1}
\eeqa
Note that we have adopted the convenient renormalization scale
of $M_K^2$ for the counterterms $L_{10}^{(0)}$ and $H_1^{(0)}$.
Of course, any other choice is also allowed, but there will be
compensating logarithmic factors as in Eq.~(\ref{R7}).

The next step is the crucial one.  One removes
simultaneously the divergent factor of ${\overline \lambda}$ and
dependence on the scale $\mu$ with the choice
\beq
2L_{10}^{(1)} + 4 H_1^{(1)} = -1 \ \ .
\label{F2}
\eeq
This result is consistent with that Ref.~\cite{gl2}, where
background field renormalization yielded the individual values
$L_{10}^{(1)}= -1/4$ and $H_1^{(1)} = -1/8$.  We are then left
with the finite and scale-independent one-loop result
\beq
\Pi_{88}(q^2)\bigg|_{\rm 1-loop} = - 2L_{10}^{(0)}(M_K^2)
- 4 H_1^{(0)}(M_K^2) - 6i{\overline B}_{21} (q^2, M_K^2) \ \ .
\label{F3}
\eeq

\begin{center}
{\bf Renormalization at Two-loop Order}
\end{center}

Renormalization at two-loop order proceeds in like manner, but
with many more terms.  Since the underlying logic is the same,
we shall not dwell on the many details of the analysis.
However, it is of some interest to observe that the leading
degree of divergence at two-loops is quadratic,
\beqa
& &\phantom{xxx}\Pi_{88}(q^2)\bigg|_{\rm 2-loop} = - \mu^{2(4-d)}
{\overline \lambda}^2
\left( {4q^2 \over F_0^2} \left[ {1\over 8} + 2K_1^{(2)} +
K_2^{(2)} \right] \right. \nonumber \\
& & \left. + {32M_K^2 \over F_0^2} \left[ K_4^{(2)} + {2\over 3}
K_3^{(2)} \right]
+ {16M_\pi^2 \over F_0^2} \left[ K_4^{(2)} - {1\over 3}  K_3^{(2)} \right]
\right) + \dots \ \ ,
\label{F4}
\eeqa
where we have exhibited only the most singular terms.  The
${\overline \lambda}^2$ divergences are removed by means of the
scale-independent choices
\beq
2K_1^{(2)} + K_2^{(2)} = -{1\over 8} \qquad {\rm and} \qquad
K_3^{(2)} = K_4^{(2)} = 0 \ \ .
\label{F5}
\eeq
Analogous procedures for the remaining part of the amplitude
turn out not to constrain the ${\cal O}(q^6)$ counterterms in
terms of numerical values as in the above equation.  Rather,
relations are obtained which connect the ${\cal O}(q^6)$ counterterms
with those from the ${\cal O}(q^4)$ sector,
\beqa
2K_1^{(1)}(\mu) + K_2^{(1)}(\mu) &=& -  L_9^{(0)}(\mu) \ , \nonumber \\
K_3^{(1)}(\mu) &=& {3\over 4}\left[ L_9^{(0)}(\mu)  +
L_{10}^{(0)}(\mu)\right]   \ ,\label{F6} \\
K_4^{(1)}(\mu) &=& {1\over 4}\left[ L_9^{(0)}(\mu)  + L_{10}^{(0)}(\mu)
\right] \ .  \nonumber
\eeqa
This completes the list of constraints.

\begin{center}
{\bf Final Form of Vector-current Propagators}
\end{center}

We express our final result for $\Pi_{88}$ in terms of
the function $i{\overline B}_{21}$ and various counterterms,
all fixed at scale $M_K^2$,
\beqa
\lefteqn{\Pi_{88}(q^2 , M_K^2) =} \nonumber \\
& & -2L_{10}^{(0)} (M_K^2) -4H_1^{(0)} (M_K^2)
-6i{\overline B}_{21}(q^2 ,M_K^2) \nonumber \\
& & + {q^2 \over F_0^2} \left[ 18 (i{\overline B}_{21}(q^2 ,M_K^2))^2
- 24 L_9^{(0)} (M_K^2)  i{\overline B}_{21}(q^2 ,M_K^2)
- P (M_K^2) \right] \nonumber \\
& & - {4M_\pi^2 \over F_0^2} \left[ R (M_K^2)
- {1\over 3} Q (M_K^2) \right]
- {8M_K^2 \over F_0^2} \left[ R (M_K^2) +
{2\over 3} Q (M_K^2) \right] \ ,  \label{F7}
\eeqa
where we define
\beqa
P(\mu) &\equiv& 4\left[ 2K_1^{(0)}(\mu) + K_2^{(0)}(\mu) +
L_9^{(-1)}(\mu) \right] \ , \nonumber \\
Q(\mu) &\equiv&
4 K_3^{(0)}(\mu) - 3 \left[ L_9^{(-1)}(\mu)  +
L_{10}^{(-1)}(\mu)\right]   \ ,\label{pqr} \\
R(\mu) &\equiv& 4K_4^{(0)}(\mu) - L_9^{(-1)}(\mu)  - L_{10}^{(-1)}(\mu)
\ .  \nonumber
\eeqa
The quantities $P$, $Q$, $R$ are the three new independent counterterms
introduced at two-loop order.  Observe that $L_9^{-1}$and $L_{10}^{-1}$
have been absorbed in these quantities.  This is an example of a
general property of the generating functional, as pointed out in
Ref.~\cite{BGS94}.

We can characterize the content of $\Pi_{88}$ as given above
in terms of `chiral counting'.  Thus the first line on the right hand
side is the one-loop result, whereas the remaining lines contain
higher-order (two-loop) contributions which are proportional to
the perturbative parameters of smallness $q^2 /F_0^2$, $M_\pi^2 /F_0^2$,
$M_K^2 /F_0^2$.  For completeness, we also write down the analogous but more
cumbersome result for $\Pi_{33}$,
\beqa
\lefteqn{\Pi_{33}(q^2 , M_\pi^2, M_K^2) =
{1\over 48 \pi^2} \ln {M_K^2\over M_\pi^2}
- {8M_K^2 \over F_0^2} R (M_K^2)}
\nonumber \\
& & -2L_{10}^{(0)} (M_K^2) -4H_1^{(0)} (M_K^2)
-4i{\overline B}_{21}(q^2 ,M_\pi^2)
-2i{\overline B}_{21}(q^2 ,M_K^2) \nonumber \\
& & + {q^2 \over F_0^2} \left[
\vphantom{\left( {1\over 16\pi^2} \ln{M_K^2 \over M_\pi^2} \right)^2}
 8 (i{\overline B}_{21}(q^2 ,M_\pi^2))^2
+ 2(i{\overline B}_{21}(q^2 ,M_K^2))^2
+ {1\over 18}\left( {1\over 16\pi^2}
\ln{M_K^2 \over M_\pi^2} \right)^2 \right. \nonumber \\
& & \left. + 8 i{\overline B}_{21}(q^2 ,M_\pi^2)i{\overline B}_{21}(q^2 ,M_K^2)
- P(M_K^2) + {L_9^{(0)} (M_K^2) \over 12\pi^2} \ln{M_K^2 \over M_\pi^2}
\right. \nonumber \\
& & \left. - \left( 2i{\overline B}_{21}(q^2 ,M_\pi^2)
+ i{\overline B}_{21}(q^2 ,M_K^2) \right)\cdot
\left( 8 L_9^{(0)} (M_K^2) + {1\over 24\pi^2} \ln{M_K^2 \over M_\pi^2}
\right) \right] \nonumber \\
& & - {4M_\pi^2 \over F_0^2} \left[ Q (M_K^2) + R (M_K^2)
+ { L_9^{(0)} (M_K^2) + L_{10}^{(0)}(M_K^2) \over 8\pi^2}\ln{M_K^2
\over M_\pi^2} \right] \ \ .
\label{F8}
\eeqa
After some algebra, one finds that the expressions in Eq.~(\ref{F7}) and
Eq.~(\ref{F8}) agree in the $SU(3)$ limit.
\section{Applications}

The vector current two-point functions may be expressed in terms
of the corresponding K\"all\'en-Lehman spectral functions.  Combined
with the high energy behaviour which follows from perturbative QCD,
the low energy representations of the two-point functions derived in
the previous section is seen to contain two pieces of information:
i) it yields the leading and next-to-leading order contributions to
the low energy expansion of the spectral functions, and ii) it implies
a set of sum rules which the spectral functions must satisfy.  We shall
discuss in turn the consequences of our results for i) and ii).

\begin{center}
{\bf Vector Spectral Functions}
\end{center}

An important application of our results involves determining the
spectral functions associated with the isospin and hypercharge
vector currents.  As pointed out in the Introduction, the former
is of special interest since it can be compared directly with
existing data.  Thus it is the isospin spectral function
$\rho_{{\rm V}33}(q^2)$ that will be discussed here.

{}From Eqs.~(\ref{P2})-(\ref{P4}), it follows that the relation
between $\Pi_{33}(q^2)$ and $\rho_{{\rm V}33}(q^2)$ is simply
\beq
\Pi_{33}(q^2) = \int_{s_0}^\infty ds\ {\rho_{{\rm V}33}(s) \over s - q^2 -
i\epsilon } \ \ ,
\label{F9}
\eeq
where $s_0 = 4M_\pi^2$ is the threshold value.  Thus the spectral function
$\rho_{{\rm V}33}(q^2)$ is proportional to the imaginary part of the
propagator,
\beq
{\rm Im}~\Pi_{33}(q^2) = \pi \theta (q^2 - 4m_\pi^2)~\rho_{{\rm V}33}(q^2) \ .
\label{F10}
\eeq
A modest amount of effort in taking the imaginary part of
Eq.~(\ref{F8}) then yields the expression
\beqa
\lefteqn{\rho_{{\rm V}33}(q^2) = {1\over 48\pi^2}
\left( 1 - {4M_\pi^2 \over q^2}
\right)^{3/2} \theta(q^2 - 4M_\pi^2)} \nonumber \\
& & \times \left[ 1 + {q^2 \over 2F_0^2} \left(
8 L_9^{(0)} (M_K^2) + {1\over 24\pi^2} \ln{M_K^2 \over M_\pi^2}
- 8 {\rm Re} i{\overline B}_{21}(q^2 ,M_\pi^2) \right. \right. \nonumber \\
& & \left. \left. - 4 {\rm Re} i{\overline B}_{21}(q^2 ,M_K^2)
\vphantom{\ln{M_K^2 \over M_\pi^2}} \right) \right]
+ {1\over 96\pi^2} \left( 1 - {4M_K^2 \over q^2}
\right)^{3/2} \theta(q^2 - 4M_K^2) \nonumber \\
& & \times \left[ 1 + {q^2 \over 2F_0^2} \left( 8 L_9^{(0)} (M_K^2)
+ {1\over 24\pi^2} \ln{M_K^2 \over M_\pi^2} \right. \right. \nonumber \\
& & \left. \left. - 8 {\rm Re} i{\overline B}_{21}(q^2 ,M_\pi^2)
- 4 {\rm Re} i{\overline B}_{21}(q^2 ,M_K^2) +
\vphantom{\ln{M_K^2 \over M_\pi^2}} \right) \right] \ .
\label{F11}
\eeqa

This result can be directly compared with data.
Since such comparison is performed near the $\pi\pi$
threshold, only the part of the above formula proportional to
$\theta(q^2 - 4M_\pi^2)$ contributes.  Some care must be exercised
in evaluating Re $i{\overline B}_{21}(q^2 ,M_\pi^2)$ and
especially Re $i{\overline B}_{21}(q^2 ,M_K^2)$.

The one-loop result was already depicted in Fig.~1.  A more complete comparison
appears in Fig.~5, where we display the individual one-loop, two-loop,
and total chiral predictions together with spectral function data.
The two-loop result given in Eq.~(\ref{F11}) contains several additive
contributions.  Of these, the largest arises from the one containing the
one-loop counterterm $L_9^{(0)} ( M_K^2)$.  For our numerical work,
we have adopted the value of $L_9^{(0)} ( M_\eta^2)$ cited in
Ref.~\cite{dgh}, with a small correction for working at scale $M_K^2$,
\beqa
L_9^{(0)} ( M_K^2) &=& L_9^{(0)} ( M_\eta^2) + {1 \over 128 \pi^2}
\log \left( {M_\eta^2 \over M_K^2}\right) \ \ , \nonumber \\
&=& 0.0071 \pm 0.0003 + 0.0002  = 0.0073 \pm 0.0003 \ .
\label{S1}
\eeqa
Another issue associated with a numerical analysis of
$\rho_{{\rm V}33}(q^2)$ is the value to be taken for the bare decay
constant $F_0$ ({\it cf}~Eq.~(\ref{F11})).  We have followed standard
procedure by employing the physical value
\beq
F_0 \quad \to \quad F_\pi \simeq 0.093~{\rm GeV} \ \ ,
\label{S2}
\eeq
since any difference between $F_0$ and $F_\pi$ will be relegated to
still higher order.  The effect of using the
one-loop result$^{\cite{gl2}}$ $F_0 < F_\pi$ would be to increase
the two-loop contribution relative to that shown in Fig.~5.
Finally, we caution that some care be taken in interpreting
a graph such as the one in Fig.~5.  For the purpose of illustration,
we have depicted the energy-dependence up to the value $s = 12~m_\pi^2$.
However, before that value is reached one expects that still
higher-loop effects should be taken into account.  In addition,
due to its stronger $s$ dependence, the two-loop `perturbation'
ultimately becomes as large as the `dominant' one-loop amplitude.
{}From Fig.~5, it would appear that our calculation is valid up
to $s \simeq 8m_\pi^2$ or an energy of about $400$ MeV.

\begin{center}
{\bf Chiral sum rules}
\end{center}

The real part of the vector current propagator depends on low energy constants
appearing at ${\cal O}(q^6)$ in the chiral expansion. Some of these constants
can be related to data by exploiting the high energy behaviour of QCD.
We are thus able to derive a set of new chiral sum rules.

The asymptotic behavior of the functions $\Pi_{aa}(q^2)$ ($a=3,8$ not
summed) is known from the operator product
expansion,$^{\cite{svz},\cite{bnp}}$
\begin{equation}
\Pi_{aa}(q^2) \sim {1\over 8\pi^2} \left(1+{\alpha_s(q^2) \over \pi}\right)
\ln{\mu^2\over -q^2} \qquad (a=3,8)\ ,
\label{assympto}
\end{equation}
where $\mu$ is the renormalization scale and $\alpha_s(q^2)$ is the QCD
running coupling constant. It follows from the asymptotic form that the
difference $\Pi_{33}-\Pi_{88}$ satisfies the unsubtracted dispersion
relation,
\begin{equation}
(\Pi_{33}-\Pi_{88})(q^2) = \int_{s_0}^\infty ds \
{(\rho_{{\rm V} 33}-\rho_{{\rm V} 88})(s) \over s-q^2-i\epsilon} \ \ .
\end{equation}
The real part of this dispersion relation can be rewritten as a set of
sum rules for the {\it negative} moments of the difference of spectral
functions,
\begin{equation}
{{\rm d}^n\over ({\rm d} q^2)^n} (\Pi_{33}-\Pi_{88}) (0)=
n! \int_{s_0}^\infty  ds \ {(\rho_{{\rm V} 33}-\rho_{{\rm V} 88})(s)\over
s^{n+1}} \qquad (n\geq 0) \ \ .
\label{Wnegn}
\end{equation}
In this form, the relevance of calculating $\Pi_{aa}(q^2)$ at
low $q^2$ is evident.

Armed with the two-loop results obtained earlier in this paper, we can
express the sum rules of Eq.~(\ref{Wnegn}) in terms of low energy
constants of the effective chiral lagrangian. For instance, we have for
the cases $n=0,1$,
\begin{eqnarray}
\lefteqn{\int_{s_0}^\infty  ds \ {(\rho_{{\rm V}33}-\rho_{{\rm V} 88})(s)
\over s} = {1\over 48 \pi^2} \ln {M_K^2\over M_\pi^2}}\nonumber\\
& &-{4 M_\pi^2\over F_0^2}\cdot {L_9^{(0)}(M_K^2)+L_{10}^{(0)}(M_K^2) \over
8 \pi^2} \ln {M_K^2\over M_\pi^2}
+{16 (M_K^2-M_\pi^2)\over 3F_0^2} Q (M_K^2)
\label{Wneg1}
\end{eqnarray}
and
\begin{eqnarray}
\lefteqn{\int_{s_0}^\infty  ds\ {(\rho_{{\rm V} 33}-\rho_{{\rm V} 88})(s)
\over s^2} = {1\over 480 \pi^2} \left({1\over M_\pi^2}-{1\over M_K^2}\right)}
\nonumber \\
& &-{1\over 18 F_0^2}\left({1\over 16 \pi^2}\right)^2
\left(3-\ln {M_K^2\over M_\pi^2}\right)\ln{M_K^2\over M_\pi^2}
+ {L_9^{(0)}(M_K^2)\over 12 \pi^2 F_0^2}\ln {M_K^2\over M_\pi^2}\ .
\label{Wneg2}
\end{eqnarray}
The sum rule in Eq.~(\ref{Wneg1}) has already been obtained by Knecht,
Moussallam and Stern.$^{\cite{KMS94}}$  However, since these authors
calculate in generalized ChPT up to ${\cal O}(q^5)$, only part of the
symmetry breaking term proportional to $Q$ appears on the RHS of their
sum rule, {\it i.e.} they necessarily lack the $L_9^{(-1)}$ and
$L_{10}^{(-1)}$ dependence occurring the defining relation for $Q$ in
Eq.~(\ref{Wneg1}).   By contrast, our expression in Eq.~(\ref{Wneg1})
is correct to ${\cal O}(q^6)$ in standard ChPT, and up to this order it
is the combination of counterterms ${4\over 3}
K_3^{(0)}-L_9^{(-1)}-L_{10}^{(-1)}$ which occurs.
We note in passing that in addition to the term proportional
to $L_9^{(0)} + L_{10}^{(0)}$, there would also be genuine two-loop
contributions if the low energy constants were evaluated at a scale
different from $M_K^2$.  Turning to the second sum rule, we see
that it contains only one low energy constant, $L_9^{(0)}$. Although
$L_9^{(0)}$ is known to rather high accuracy from the electromagnetic
charge radius of the pion,$^{\cite{gl2}}$ it is nevertheless true that
the relation in Eq.~(\ref{Wneg2}) offers a nontrivial check on the
determination of this quantity.

Our two-loop representation of $\Pi_{33}-\Pi_{88}$ allows one
to calculate the leading and next-to-leading order contributions to
the RHS for all sum rules defined by Eq.~(\ref{Wnegn}) if $n\geq 0$.
The leading term comes from the one-loop expression and is determined
entirely by the $\pi\pi$ and $K{\bar K}$ continuum terms.  The next-to-leading
order term gets contributions from both continuua and resonances, the
latter being parametrized by low energy constants like $L_9^{(0)}$.
It might well be that these contributions are numerically as important
as the leading term since the relevant physical phenomenon enters the
calculation first only at that level of approximation.

Thus far, we have considered only the set of chiral sum rules
associated with the difference, $\Pi_{33}-\Pi_{88}$, of the
polarization functions.  We can also derive sum rules for the individual
vacuum polarizations.  According to the asymptotic behaviour given in
Eq.~(\ref{assympto}), at least one subtraction is required to obtain
convergent sum rules.  Thus we can write
\begin{equation}
\int_{s_0}^\infty  ds \ {\rho_{{\rm V} aa}(s) \over s^{n+1}}=
{{\rm d}^n\over ({\rm d} q^2)^n} \Pi_{aa} (0) \qquad (a=3,8)
\end{equation}
for $n\geq 1$.  As an example, we give the explicit expression for
the isospin vacuum polarization $a=3$ with $n=1$,
\begin{eqnarray}
\lefteqn{\int_{s_0}^\infty  ds \ {\rho_{{\rm V} 33}(s) \over s^2}
= {1\over 480 \pi^2} \left({1\over M_\pi^2}+{1\over 2 M_K^2}\right)}\nonumber\\
& &+{1\over 8 F_0^2}\left({1\over 16 \pi^2}\right)^2
\left(1-{2\over 3}\ln {M_K^2\over M_\pi^2}\right)^2
- {L_9^{(0)}(M_K^2)\over 8 \pi^2 F_0^2}
\left(1-{2\over 3}\ln {M_K^2\over M_\pi^2}\right)\nonumber\\
& &-{1\over F_0^2} P (M_K^2) \ .
\label{W3neg2}
\end{eqnarray}
This sum rule clearly serves to fix the ${\cal O}(q^6)$ renormalization
constant $P$.  Note that the $s^{-n}$ sum rules which involve just the
individual spectral functions are more slowly convergent than those
involving a flavor difference of spectral functions because for the
latter the leading term in the asymptotic expansion cancels.

To analyze the sum rules of Eqs.~(\ref{Wneg1})--(\ref{W3neg2}), it will be
necessary to evaluate integrals containing negative moments of the
spectral functions.  We shall relegate such studies to a forthcoming
publication, where we will also attempt to give realistic error bars
on our determinations.  Here, it is sufficient to touch upon the
general strategy for calculating such integrals.  As in Ref.~\cite{dg1}, it
is natural to divide the region of integration into three parts: A) the
threshold region, B) an intermediate energy region where experimental
information on the spectral functions is available, and C) the asymptotic
region. The asymptotic behaviour of the spectral functions is calculated
using the OPE. As such, we anticipate the contribution from Region C to be
especially small for the flavor-difference sum rules since for these the
OPE will be proportional to light quark masses.  Region B is the most
difficult one to treat because it is here that intricacies of data
analysis enter.  Finally, in Region A we can use the improved
low energy expansion of the spectral functions obtained in this paper.

What will be the outcome of numerically analyzing the new chiral rules?
The answer is that one thereby determines the combinations of counterterms,
particularly those of ${\cal O}(q^6)$, appearing in them.  By no means is
this a trivial task, as the following example will demonstrate. For
definiteness, we focus on the sum rule of Eq.~(\ref{Wneg1}) and rewrite it as
\begin{eqnarray}
\lefteqn{\int_{s_0}^\infty  ds \ {(\rho_{{\rm V}33}-\rho_{{\rm V} 88})(s)
\over s} \ -\  {1\over 48 \pi^2} \ln {M_K^2\over M_\pi^2} \ =}\nonumber\\
& & -{4 M_\pi^2\over F_0^2}\cdot {L_9^{(0)}(M_K^2)+L_{10}^{(0)}(M_K^2)
\over 8 \pi^2} \ln {M_K^2\over M_\pi^2} + {16 (M_K^2-M_\pi^2)\over
3F_0^2} Q(M_K^2)
\label{mod}
\end{eqnarray}
so that only terms proportional to the ${\cal O}(q^4)$ and ${\cal O}(q^6)$
low energy constants appear on the RHS.  We see from
Eq.~(\ref{mod}) that these counterterms are fixed by the difference
between a spectral integral and a logarithmic mass ratio.  However, the
latter term can itself be expressed as a kind of spectral integral by
employing the leading-order\footnote{Similar arguments apply to
next-to-leading order, but the corresponding expressions are much more
complicated.} chiral representations for the isospin and hypercharge
spectral functions.  To demonstrate this, let us first define energy
scales $\Lambda_3$ and $\Lambda_8$ below which the leading-order chiral
representations $\rho^{\rm (chir)}_{{\rm V} 33}$ and
$\rho^{\rm (chir)}_{{\rm V} 88}$ are reasonable approximations to the
physical spectral functions.  We then carry out the integration to obtain
\begin{equation}
\int^{\Lambda_3}_{4M_\pi^2} ds~{\rho^{\rm (chir)}_{{\rm V} 33}(s)\over s} -
\int^{\Lambda_8}_{4M_K^2} ds~{\rho^{\rm (chir)}_{{\rm V} 88}(s)\over s}
\doteq {1\over 48 \pi^2} \left[ I\left({4 M_\pi^2 \over \Lambda_3}\right)
- I\left({4 M_K^2 \over \Lambda_8} \right)\right] \ ,
\label{threshold}
\end{equation}
where the function $I(s_0 /\Lambda)$ is given by
\begin{eqnarray}
\lefteqn{I\left({s_0 \over \Lambda} \right) \equiv \int_{s_0}^\Lambda
{ds \over s} \left[ 1- {s_0 \over s}\right]^{3/2}} \label{expl} \\
& &= -{2\over 3} \left[ 1-{s_0\over \Lambda}\right]^{3/2}
- 2 \left[ 1 - {s_0\over \Lambda}\right]^{1/2}
-\ln{s_0\over \Lambda}
+ 2 \ln\left[ 1+\left(1-{s_0\over \Lambda}\right)^{1/2} \right]~.
 \nonumber
\end{eqnarray}
Upon removing the cutoffs in an appropriate manner, one regains the
logarithmic term appearing on the LHS in Eq.~(\ref{mod}), {\it viz.}
\beq
\lim_{ \Lambda_{3,8} \to \infty \atop \Lambda_3 \to\Lambda_8 }
\left[ I\left({4 M_\pi^2 \over \Lambda_3}\right)
- I\left({4 M_K^2 \over \Lambda_8} \right)\right]
= \ln\left({M_K^2 \over M_\pi^2}\right) \ .
\label{lmt}
\eeq
The point of this exercise is to stress that determination
of the counterterms will involve spectral integrals over $\Lambda
\le s < \infty$ of the difference $\rho_{\rm V}^{\rm (phys)} -
\rho_{\rm V}^{\rm (chir)}$.  The reader should appreciate the
irony that, although negative moments enhance the threshold region
in the new chiral sum rules, this has little influence on the counterterm
determination.  That is, as just shown the threshold region, in which
$\rho_{\rm V}^{\rm (phys)} \simeq \rho_{\rm V}^{\rm (chir)}$, is
suppressed via cancelation.  Thus, to correctly infer values for the
counterterms one must carefully analyze regions in which $\rho_{\rm V}^{\rm
(phys)} \ne \rho_{\rm V}^{\rm (chir)}$, most crucially the intermediate
energy region where available data fix the spectral functions.

Finally, let us point out that our discussion of Eq.~(\ref{mod}) has
yielded yet another insight.  If one passes to the limit of, say,
$SU(2)$ chiral symmetry, then the explicit logarithmic dependence on
$M_\pi^2$ in Eq.~(\ref{mod}) leads to a divergence.  In order to maintain
consistency, then so too must the spectral integral in Eq.~(\ref{mod})
diverge.  The discussion just given shows precisely, via the chiral
representations for the spectral functions, how this comes about.
Thus the explicit logarithmic dependence in the chiral sum rules correctly
reflects the leading implicit infrared singularities of the spectral
integrals which appear therein.

\section{Conclusions}

We have used ChPT to determine the two vector current propagators
$\Pi_{33}$ and $\Pi_{88}$ through two-loop order.  The main
results appear in Eq.~(\ref{F7}) and Eq.~(\ref{F8}).  As expected,
one finds the analytic structure corresponding to both $\pi\pi$ and
$K{\bar K}$ thresholds for $\Pi_{33}$ and just the
$K{\bar K}$ threshold for $\Pi_{88}$.  There is no
$3\pi$ threshold for the propagators because we do not consider anomalous
currents in this paper.  The final results are seen to contain a total of
six renormalization constants and this warrants some
discussion.$^{\cite{rem}}$  The presence of the ${\cal O}(q^4)$ counterterms
$L_{10}^{(0)}$ and $H_{1}^{(0)}$ already follows from the original one-loop
analysis of Gasser and Leutwyler, whereas the remaining four arise from
two-loop effects.  Although the quantity $L_{9}^{(0)}$ is known from the
one-loop analysis of the pion form factor, the other three
($P$, $Q$, $R$) are new.  As we have seen, all
but $L_{9}^{(0)}$ contribute only to the real parts of the
isospin and hypercharge propagators.
The presence of these new renormalization constants is, of course, no
surprise.  For an effective theory like ChPT, the number of terms in
the lagrangian increases sharply with loop order, and the renormalization
constants are simply the set of quantities needed to render the theory
meaningful at a given order.  We shall comment shortly on how to
partially constrain the new renormalization constants.

There are a number of interesting applications of the work described
here.  Of these, the most immediate involves the comparison between
data and the chiral prediction for the isospin spectral function
$\rho_{{\rm V}33}(q^2)$. Our calculation allows extraction of the
isospin and hypercharge spectral functions at next-to-leading order in
the chiral expansion. For $\rho_{{\rm V}33}(q^2)$ the correction turns out
to be substantial even at energies very close to the two-pion
threshold, {\it cf.} Fig.~5. From this figure we can see that our
two-loop representation of the isospin spectral function is valid
up to an energy of about 400 MeV.

Mention was made above of the counterterms which contribute to
our final results.  Three of these are new in the sense that
they lie outside the phenomenology already established for the
one-loop sector.  They all occur in the real parts of the
isospin and hypercharge propagators. By exploiting the asymptotic
behaviour of the vector polarization functions $\Pi_{33}$ and $\Pi_{88}$
as inferred from the operator product expansion, we have derived a set
of new chiral sum rules.  Among these is a sum rule for the ${\cal O}(q^4)$
counterterm $L_9^{(0)}$. Since this quantity is already well-known, the
sum rule offers a nontrivial check on the theory. The other sum rules fix
${\cal O}(q^6)$ counterterms, some of which also appear in the two-loop
analysis of the process $\gamma\gamma\rightarrow \pi^0\pi^0$. We have
begun the program of study outlined in Sect.~7 to constrain these
quantities in terms of data, and results will be announced elsewhere.

Finally, work is well underway on a two-loop determination of
the $SU(3)$ isospin and hypercharge axialvector current propagators.
This study is a natural extension of both the present one
and also of previous work done on the generalized Weinberg sum
rules.$^{\cite{dg1}}$  Although a chiral analysis of the
axialvector propagators will superficially resemble that of the vector
propagators, the two turn out to be very different in practice.
In particular, a great deal of work on the renormalization of
masses and decay constants of pseudoscalar mesons is required for
the latter.  Thus, we shall treat the axialvector sector separately in a
forthcoming publication.

\begin{center}
{\bf Acknowledgements}
\end{center}

The research described in this paper was supported in part by
the National Science Foundation and by Schweizerischer Nationalfonds.
One of us (E.G.) wishes to acknowledge the kind hospitality of the
theory group at the Rutherford Appleton Laboratory, where part of the
work described herein was performed.

\vfill
\eject
\appendix{\bf Appendix A: Feynman Integrals}

In this Appendix, we shall collect together definitions and
properties of functions which play a
central role in our analysis.  In each case, one begins with an
integral defined in $d$-dimensions and proceeds with its evaluation
by isolating any divergence it might contain and determining the
associated finite part.

\begin{center}
{\bf The Integral A}
\end{center}

The simplest such quantity is the scalar integral
\beq
A ( m^2 ) \equiv \int {d^d k\over (2\pi)^d}
 ~ { 1 \over k^2 - m^2 } \ \ .
\label{a1}
\eeq
The evaluation procedure is standard and we obtain
\beqa
A ( m^2 ) &=& {-i \over (4 \pi)^{d/2} }
{\mu^{4-d} \over \mu^{4-d}} \Gamma\left( 1 - {d\over 2} \right)
(m^2 )^{(d - 2 )/2} \label{a2} \\
&=& \mu^{d-4} \left[ -2i m^2 {\overline \lambda}
- { im^2 \over 16 \pi^2} \log \left( {m^2 \over \mu^2} \right)
+ \dots \right] \ \ ,
\label{a3}
\eeqa
where
\beq
\lambda = \mu^{d - 4} {\overline \lambda} =
{ \mu^{d - 4} \over 16 \pi^2} \left[ {1 \over d - 4} - {1\over 2}
\left( \log {4\pi} - \gamma + 1 \right) \right] \ \ .
\label{a4}
\eeq
The quantity $\mu$, introduced in Eq.~(\ref{a2}), is the mass scale
which enters the calculation via the use of dimensional regularization.
The $\mu^{d - 4}$ prefactor in Eq.~(\ref{a3}) ensures that $A ( m^2 )$
has the proper units in $d$-dimensions.

\begin{center}
{\bf The Integrals B, B$_\mu$, B$_{\mu\nu}$}
\end{center}

Next we define
\beqa
B_{\mu\nu}(q^2 , m^2) &=& \int {d^d k\over (2\pi)^d}
 ~ {k_\mu k_\nu \over (k^2 - m^2) \cdot ((q - k)^2 - m^2)} \ ,
\label{a5} \\
B_\mu(q^2, m^2)  &=& \int {d^d k\over (2\pi)^d}
 ~ {k_\mu \over (k^2 - m^2) \cdot ((q - k)^2 - m^2)} \ ,
\label{a6} \\
B(q^2 ,m^2) &=& \int {d^d k\over (2\pi)^d}
 ~ {1 \over (k^2 - m^2) \cdot ((q - k)^2 - m^2)}
\ \ . \label{a7}
\eeqa
Since the scalar integral $B$ is divergent, it is often convenient to
work with a function ${\overline B}$ made finite by subtraction,
\beq
{\overline B}(q^2 ,m^2)  \equiv B(q^2 ,m^2) - B(0 ,m^2) \ \ ,
\label{a7a}
\eeq
where, with the aid of Eq.~(\ref{a3}), we have
\beq
B (0 ,m^2) =  {\partial A (m^2) \over \partial m^2}
= { A(m^2 ) \over m^2} - {i \over 16 \pi^2} \ \ .
\label{a7b}
\eeq
The subtracted integral can be written as
\beqa
{\overline B}(q^2 ,m^2) &=& - {i \over 16\pi^2} \int_0^1 dx\
\log\left( 1 - x(1-x) {q^2 \over m^2} \right) \nonumber \\
&=& {i\over 96\pi^2}\cdot {q^2 \over m^2}
+ {i\over 960\pi^2}\cdot {q^4 \over m^4} + \dots \ \ .
\label{a7c}
\eeqa
To study $B_\mu$, we invoke covariance to write $B_\mu(q^2, m^2)
\equiv B_1 q_\mu$.
Then, by considering $q^\mu B_\mu$ one finds $B_1 = B/2$ so that
\beq
B_\mu(q^2, m^2)  = {q_\mu \over 2} B(q^2 ,m^2) \ \ .
\label{a8}
\eeq
The tensor integral $B_{\mu\nu}$ can be conveniently decomposed as
\beq
B_{\mu\nu}(q^2 , m^2) \equiv B_{21}(q^2 , m^2)  q_\mu q_\nu  +
B_{22}(q^2 , m^2)  g_{\mu\nu} \ \ .
\label{a9}
\eeq
For notational simplicity, we shall temporarily
suppress function arguments in the following.
Study of the contraction $q^\mu B_{\mu\nu}$ leads to the relation
\beq
q^2 B_{21} + B_{22} = {1\over 2} A + {q^2 \over 4} B \ \ ,
\label{a10}
\eeq
whereas the trace $g^{\mu\nu}B_{\mu\nu}$ yields
\beq
q^2 B_{21} + d~B_{22} = A + m^2 B \ \ .
\label{a11}
\eeq
{}From these relations, it follows that
\beqa
B_{21} &=& {1\over 3} \left[ \left( 1 - {m^2 \over q^2} \right)
{\overline B} + {A \over m^2} - {5i \over 96 \pi^2} \right] \nonumber \\
B_{22} &=& -{q^2\over 12} \left[ \left( 1 - {4m^2 \over q^2} \right)
{\overline B} + {A \over m^2} \left( 1 - {6m^2 \over q^2} \right)
- {i \over 48 \pi^2} \right] \ \ .
\label{a12}
\eeqa

A combination of these integrals which appears naturally in loop
diagrams is
\beqa
T_{\mu\nu} &\equiv& 4 B_{\mu\nu} + q_\mu q_\nu B - 2q_\mu B_\nu
- 2q_\nu B_\mu \nonumber \\
&=& 4 B_{\mu\nu} - q_\mu q_\nu B \label{a13} \\
&=& (q_\mu q_\nu - g_{\mu\nu} q^2 ) ( 4 B_{21} - B ) +
g_{\mu\nu} ( 4 B_{22} + 4q^2 B_{21} - q^2 B ) \ , \nonumber
\eeqa
where we have introduced a convenient tensor decomposition for $T_{\mu\nu}$
in the last line.  The relations in Eq.~(\ref{a12}) can then be used to obtain
\beq
T_{\mu\nu}  = (q_\mu q_\nu - g_{\mu\nu} q^2 )
\left( 4 {\overline B}_{21} + {A \over 3m^2}\right) + 2A g_{\mu\nu} \ ,
\label{a13a}
\eeq
where we define the finite quantity ${\overline B}_{21}$ as
\beq
{\overline B}_{21} (q^2 , m^2 ) \equiv {1 \over 12} \left[
\left( 1 - {4m^2 \over q^2} \right) {\overline B} -
{i \over 48 \pi^2} \right] \ \ .
\label{a14}
\eeq

\begin{center}
{\bf The Integrals C, C$_\mu$, C$_{\mu\nu}$}
\end{center}

Finally we consider the set of integrals
\beqa
C_{\mu\nu}(q^2 , m^2) &=& \int {d^d k\over (2\pi)^d}
 ~ {k_\mu k_\nu \over (k^2 - m^2)^2 \cdot ((q - k)^2 - m^2)} \ ,
\label{a15} \\
C_\mu(q^2, m^2)  &=& \int {d^d k\over (2\pi)^d}
 ~ {k_\mu \over (k^2 - m^2)^2 \cdot ((q - k)^2 - m^2)} \ ,
\label{a16} \\
C (q^2 ,m^2) &=& \int {d^d k\over (2\pi)^d}
 ~ {1 \over (k^2 - m^2)^2 \cdot ((q - k)^2 - m^2)}
\ \ . \label{a17}
\eeqa
We first use covariance to write the vector integral as
$C_\mu = C_1 q_\mu$.  Analysis of the contracted form
$q_\mu C^\mu$ then determines $C_1$ in terms of $C$,
\beq
C_1 (q^2 ,m^2) = {1 \over 2q^2} \left[ q^2 C (q^2 ,m^2)
+ {\overline B}(q^2 ,m^2)  \right] \ \ .
\label{a18}
\eeq
The scalar integral $C$ can itself be determined by beginning with
the identity $q_\mu \partial B /\partial q_\mu =
-2 q_\mu C^\mu$, from which it follows
\beq
C (q^2 ,m^2)  = {1 \over q^2 - 4m^2 } \left( -{\overline B}(q^2 ,m^2)
+ {i \over 8\pi^2} \right) \ \ .
\label{a19}
\eeq
Analogous to our treatment of the tensor integral $B_{\mu\nu}$, we write
the quantity $C_{\mu\nu}$ as
\beq
C_{\mu\nu}(q^2 , m^2) \equiv C_{21}(q^2 , m^2)  q_\mu q_\nu  +
C_{22}(q^2 , m^2)  g_{\mu\nu} \ \ .
\label{a20}
\eeq
As before, the contraction $q^\mu C_{\mu\nu}$ leads to one relation,
\beq
q^2 C_{21} + C_{22} = {1\over 4} \left[ q^2 C + B + {\overline B} \right] \ \ ,
\label{a21}
\eeq
and the trace $g^{\mu\nu} C_{\mu\nu}$ yields another,
\beq
q^2 C_{21} + d~C_{22} = B + m^2 C \ \ .
\label{a22}
\eeq
Finally we define a quantity $U_{\mu\nu}$ which, like $T_{\mu\nu}$,
occurs in loop diagrams,
\beqa
U_{\mu\nu} &\equiv& 4 C_{\mu\nu} + q_\mu q_\nu C - 2q_\mu C_\nu
- 2q_\nu C_\mu \nonumber \\
&=& (q_\mu q_\nu - g_{\mu\nu} q^2 ) \left( 4 C_{21} - C - 2{ {\overline B}
\over q^2} \right) \nonumber \\
& & \phantom{xxxx} + g_{\mu\nu} \left( 4 C_{22} + q^2 \left[ 4 C_{21} - C
- {2 {\overline B} \over q^2} \right] \right) \ .
\label{a23}
\eeqa
With the input of the relations in Eqs.~(\ref{a21}),(\ref{a22}), the
final line in Eq.~(\ref{a23}) results in the useful expression
\beq
U_{\mu\nu} = -(q_\mu q_\nu - g_{\mu\nu} q^2 ) {{\overline B} \over q^2}
+ g_{\mu\nu} \left( {A \over m^2} - {i \over 16 \pi^2} \right) \ \ .
\label{a24}
\eeq
\appendix{\bf Appendix B: The Isospin Vector Spectral Function}

In the following, we shall show how to obtain an expression
for the vector-current spectral function good to two-loop
order by using knowledge of the one-loop formulae for the
pseudoscalar meson form factors.  This will serve as a useful
check of our two-loop analysis.  For definiteness, we shall
focus on the isospin spectral function.

We begin with the statement of unitarity for the isospin
vector-current propagator.  At sufficiently low energies, only
the physical two-pion intermediate state contributes.  We can use
Eq.~(15) of Ref.~\cite{dg1} and surrounding discussion to write
\beq
\rho_{{\rm V}33}(s) = {1 \over 48 \pi^2}
\left[ 1 - {4M_\pi^2\over s} \right]^{3/2} |F_{\pi\pi}(s)|^2 \ \ .
\label{b1}
\eeq
If we interpret both sides of this equation in the context of a chiral
expansion, then the spectral function to $(n+1)^{st}$ order is expressible
in terms of the pion form factor given to $n^{th}$ order. For example, in
this way one immediately recovers the one-loop isospin spectral function
of Ref.~\cite{gl1},
\beq
F_{\pi\pi}(s)\bigg|_{\rm tree} = 1 \qquad \Longrightarrow \qquad
\rho_{{\rm V}33}(s)\bigg|_{\rm 1-loop}
= {1 \over 48 \pi^2} \left[ 1 - {4M_\pi^2\over s} \right]^{3/2} \ .
\label{b2}
\eeq

The corresponding prediction for the spectral function at two-loop
order is obtained in like manner, although requiring more work.
The chiral expansion of the pion form factor appears to first
order in Ref.~\cite{glff},
\beq
F_{\pi\pi}(s) = 1 + 2 H(s, M_\pi^2) + H(s, M_K^2) +
\dots \ \ ,
\label{b3}
\eeq
where, in terms of our notation,
\beq
H(s, M^2) \equiv {2s \over 3F_0^2} ~ L_9^{(0)} ( M^2) -
{s \over F_0^2} ~i{\overline B}_{21}(s, M^2) \ \ .
\label{b4}
\eeq
It then follows to two-loop order that
\beqa
\rho_{{\rm V}33}(s) &=& {1 \over 48 \pi^2} \left[ 1 - {4M_\pi^2\over s}
\right]^{3/2} \nonumber \\
& & \times \left( 1 + 4 {\rm Re} H(s, M_\pi^2 ) + 2 {\rm Re}
H(s, M_K^2 ) \right) + \dots \nonumber \\
&=& {1 \over 48 \pi^2} \left[ 1 - {4M_\pi^2\over s}
\right]^{3/2} \left( 1 + {2s \over 3F_0^2} \left[ 4L_9^{(0)}  ( M_\pi^2) +
2L_9^{(0)} ( M_K^2)\right] \right.\nonumber \\
& &\left.  - {s \over F_0^2} \left[ 4i{\overline B}_{21}(s, M_\pi^2) +
2i{\overline B}_{21}(s, M_K^2) \right] \right) \ \ .
\label{b6}
\eeqa
Upon noting that
\beq
L_9^{(0)} ( M_\pi^2) = L_9^{(0)}  ( M_K^2) + {1 \over 128 \pi^2}
\log \left( {M_K^2 \over M_\pi^2}\right) \ \ ,
\label{b7}
\eeq
we obtain precisely the two-loop result of Eq.~(\ref{F11}) derived
earlier in this paper,
\beqa
\lefteqn{\rho_{{\rm V}33}(s) = {1 \over 48 \pi^2} \left[ 1 -
{4M_\pi^2\over s} \right]^{3/2}
\Bigg[ 1 + {s \over F_0^2}\Big[ 4 L_9^{(0)} ( M_K^2)} \nonumber \\
& & + {1 \over 48 \pi^2} \log \left(
{M_K^2 \over M_\pi^2}\right) - 4i{\overline B}_{21}(s, M_\pi^2) -
2i{\overline B}_{21}(s, M_K^2) \Big]\Bigg]\ . \label{b8}
\eeqa
\begin{center}{\bf\large Figure Captions}
\end{center}
\vspace{0.3cm}
\begin{flushleft}
Fig.~1 \hspace{0.2cm} Comparison of 1-loop chiral prediction with data. \\
\vspace{0.3cm}
Fig.~2 \hspace{0.2cm} Generic graph for computing vector-current
propagators. \\
\vspace{0.3cm}
Fig.~3 \hspace{0.2cm} One-loop graphs for $\Delta_{{\rm V}~ab}^{\mu\nu}$. \\
\vspace{0.3cm}
Fig.~4 \hspace{0.2cm} Two-loop graphs for $\Delta_{{\rm V}~ab}^{\mu\nu}$. \\
\vspace{0.3cm}
Fig.~5 \hspace{0.2cm} Comparison of 2-loop chiral prediction with data.
The one-loop, two-loop and total contributions are denoted respectively
by dotted, dashed and solid lines. \\
\vspace{0.3cm}
\end{flushleft}
\vfill \eject

\begin{thebibliography}{99}
\bibitem{ff} We shall show in Appendix~B that one can also use the
one-loop pion form factor to obtain this information.  Thus one
has an independent check for the correctness of our two-loop analysis.
\bibitem{spec} We work with covariantly defined T-products.
\bibitem{rem} We remind the reader that we have chosen to fix
all such constants at a common renormalization energy scale, $M_K^2$.
\end{thebibliography}
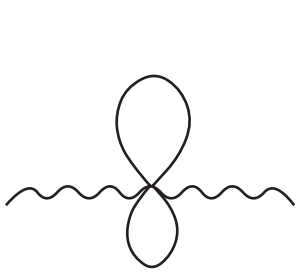
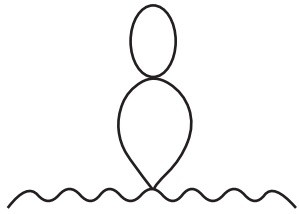
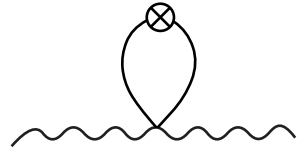
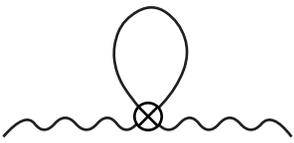
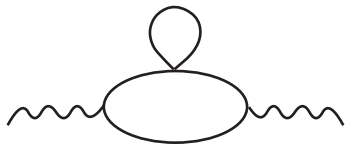
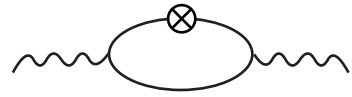
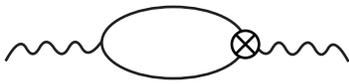
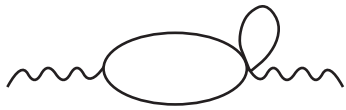
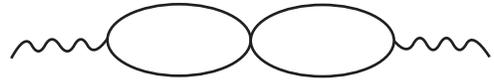
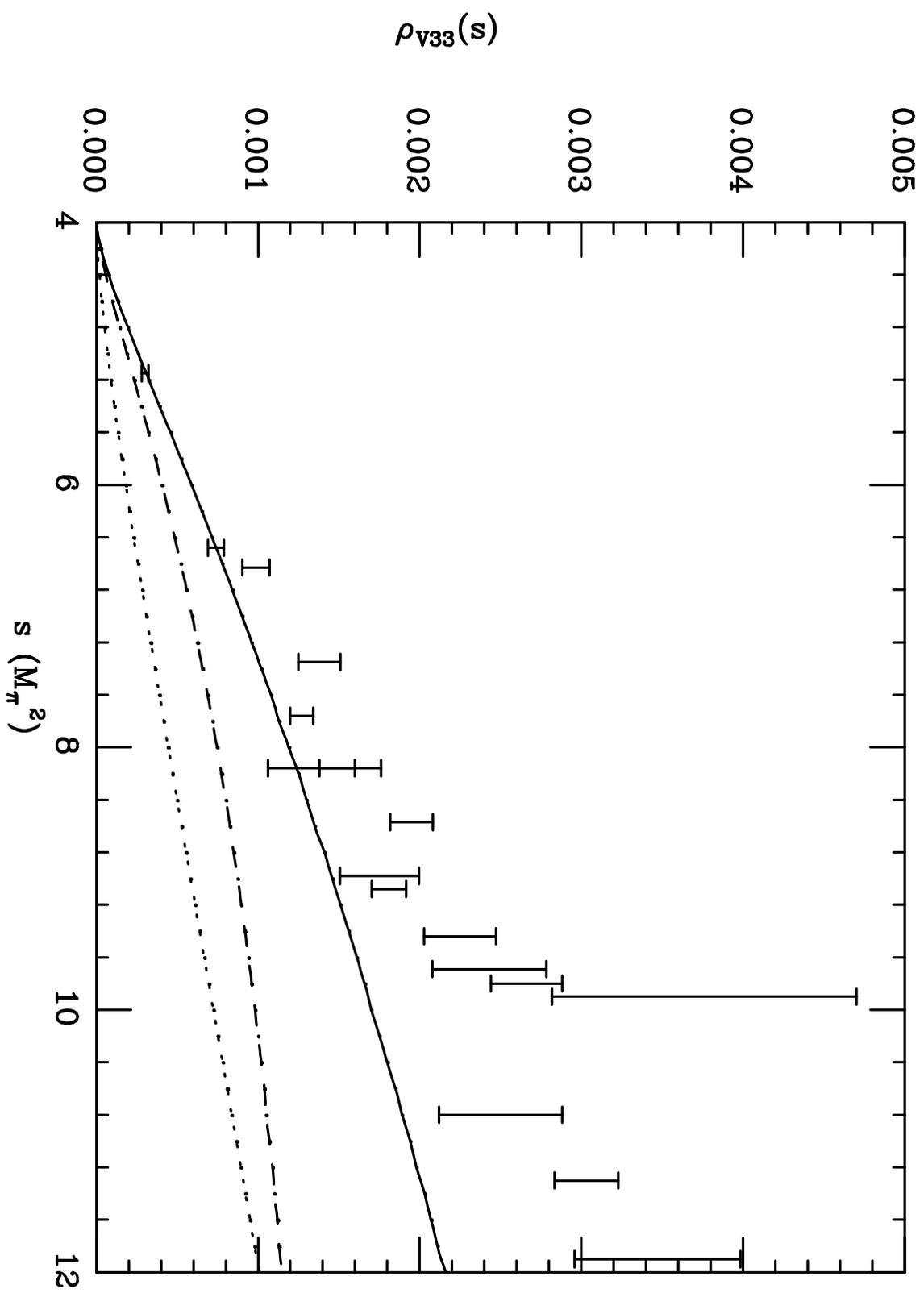
\end{document}